\documentclass{aastex63}
\shorttitle{Bayesian Characterization of Binaries in NGC 188} 
\shortauthors{Cohen, Geller \& von Hippel} 
\begin{document}
\title{Bayesian Characterization of Main Sequence Binaries in the Old Open Cluster NGC 188}

\correspondingauthor{Roger E. Cohen}
\email{rcohen@stsci.edu}

\author{Roger E. Cohen}
\affiliation{Space Telescope Science Institute, 3700 San Martin Drive, Baltimore, MD, USA}

\author{Aaron M. Geller}
\affiliation{Center for Interdisciplinary Exploration and Research in Astrophysics (CIERA) and Department of Physics \& Astronomy, Northwestern University, 2145 Sheridan Rd., Evanston, IL 60208, USA}
\affiliation{Department of Astronomy, Adler Planetarium,  1300 S. Lake Shore Drive, Chicago, IL 60605, USA} 

\author{Ted von Hippel}
\affiliation{Department of Physical Sciences, Embry-Riddle Aeronautical University, Daytona Beach, FL, USA}

\begin{abstract}

The binary fractions of open and globular clusters yield powerful constraints on their dynamical
state and evolutionary history.  We apply publicly available Bayesian analysis tools to
a $UBVRIJHK_{S}$ photometric catalog of the open cluster NGC 188 to detect and characterize
photometric binaries along the cluster main sequence.  This technique has the advantage that
it self-consistently handles photometric errors, missing data in various bandpasses, and 
star-by-star prior constraints on cluster membership.  
Simulations
are used to verify uncertainties and quantify selection biases in our analysis, illustrating that among binaries with mass ratios $>$0.5, we recover the binary fraction to better than 7\% in the mean, with no significant dependence on binary fraction and a mild dependence on assumed mass ratio distribution.  
Using our photometric catalog, we recover the majority (65\%$\pm$11\%) of spectroscopically identified main sequence binaries, including 8 of the 9 with spectroscopically measured mass ratios.  Accounting for incompleteness and systematics, we derive a mass-ratio distribution that rises toward lower mass ratios (within our $q>0.5$ analysis domain).  We observe a raw binary fraction for solar-type main sequence stars with mass ratios $q >$0.5 of 42\%$\pm$4\%, independent of the assumed mass ratio distribution to within its uncertainties, consistent with literature values for old open clusters but significantly higher than the field solar-type binary fraction. 
We confirm that the binaries identified by our method are more concentrated than single stars, in agreement with previous studies, and we demonstrate that the binary nature of those candidates which remain unidentified spectroscopically is strongly supported by photometry from \textit{Gaia} DR2.
\end{abstract}


\section{Introduction} \label{sec:intro}
\subsection{The Utility of Binary Properties}

Binary stars influence the dynamical evolution of star clusters, and in turn are observational tracers of the dynamical states and histories of present-day star clusters.  The primary observable properties of a binary population include the binary fraction, and distributions of orbital periods, eccentricities and mass ratios.    
With increasingly sophisticated modelling tools, we can now aim to match very detailed observed characteristics of real star clusters (including their binaries) within simulations, in hopes of revealing the clusters' histories, with applications including the origin of stellar exotica such as blue stragglers  \citep{gellernbody} as well as the nature and evolution of molecular clouds from which clusters form \citep{spin}. 
The ability for such models to provide accurate predictions relies on their ability to match the observed binary population.

Star cluster models make numerous predictions for the effects of dynamics on a binary population, over many Gyr.  In this paper, we will primarily focus on the binary fraction and mass-ratio distribution.  The binary fraction can be modified over time by both distant two-body relaxation effects, and close encounters with other stars and binaries.  For instance, the effects of two-body relaxation leading to mass segregation should raise the main-sequence binary fraction in the cluster core relative to the cluster halo, as the binaries are generally more massive than the single stars (at least within a magnitude limited sample). The degree to which the binaries are mass segregated should increase with the number of relaxation times the cluster has lived.  Close stellar encounters will truncate the distribution of orbital separations (or periods) at the "hard-soft" boundary \citep[e.g.][]{heggie75, gellernbody}.  For clusters with larger velocity dispersions, this hard-soft boundary moves in toward tighter binaries (and shorter periods).  One consequence of this could be to lower the binary fraction (due to the disruption of wide systems) for higher-mass clusters (which generally have higher velocity dispersions).  

Encounters are also expected to modify the mass-ratio distribution of a binary population.  For instance, a common outcome of an encounter between a binary and a single star, where the least massive star initially resides within the binary, is to exchange the more massive single star into the binary, and leave the least massive star as a single.  Through such exchange encounters, the mass-ratio distribution of the binary population can become biased toward higher mass ratios (a mass-ratio of unity, $q=1$, is an equal-mass binary).  

Thus, observing the binary fraction and the mass-ratio distribution of a population of binaries in a star cluster can provide valuable insights into the cluster dynamical environment.  Of particular importance will be to compare these properties, derived in a standard way, across different star clusters (and also to similar observed properties in the Galactic field).  Furthermore, observed binary properties in young star clusters that have not experienced very significant dynamical evolution are essential for guiding the initial conditions of star cluster models, while the binary properties of older, dynamically evolved clusters may be useful to constrain their dynamical histories.  

Traditionally, long-term spectroscopic surveys of star clusters have been employed to derive individual binary orbits, and then construct binary distributions \citep{gellerorbits, gellerhardbin, milliman_6819, leiner_m35, geller_m67}.  When spectroscopic information is unavailable to constrain the presence of binary stars or their properties, various approaches have been taken to characterize cluster binary populations photometrically.  For well-populated Galactic globular clusters, \citet{milone} used exquisite two-filter \textit{Hubble Space Telescope} imaging and artificial star tests to measure the binary fraction in 59 Galactic globular clusters (GGCs).  Their analysis focused on higher mass ratio ($q>0.5$) binaries which are most clearly characterized photometrically, finding that GGCs tend to have low ($\lesssim$10\%) binary fractions, and this fraction correlates with radius within a cluster, and also correlates with other cluster properties among the GGCs in their sample.  Regarding open clusters, \citet{binocs} fit spectral energy distributions (SEDs) of cluster members over a wide range of broadband filters (near-UV to mid-IR) to characterize the binary fractions of open clusters.  Using a sample of 8 open clusters, they report binary fractions of $\sim$40\% for clusters with ages of 1 Gyr or older, with a marked increase to $>$60\% for clusters younger than $\sim$200 Myr.  However, without a larger sample, it is unclear how this trend depends on other cluster properties (metallicity, environment, dynamical state), and it is also unclear to what extent this methodology is sensitive to assumed cluster parameters such as distance and reddening.  

While a Bayesian approach has been used in the past to constrain the properties of multiple systems \citep{widmark}, here we use a methodology (described in Sect.~\ref{bayessect}) which simultaneously solves for cluster distance, reddening, age and metallicity, but exploits star-by-star information both as optional input priors (on cluster membership) and as outputs (via primary and secondary mass distributions for each star).  To validate this methodology, here we apply it to an open cluster with an extensive database of extant binary parameters obtained through long-term spectroscopic campaigns. 

\subsection{NGC 188 as a Test Case}

Here, we study in detail the old ($\sim$7 Gyr) solar-metallicity \citep{ata, friel, shane} open cluster NGC 188, which is one of the very few extensively studied, old rich open clusters.  We refer the reader to \citet{fornal} and \citet{gellerrvs} and references therein for an overview of the previous observational surveys for the cluster.  Most importantly for this study, NGC 188 has multi-band photometry ranging from the infrared through the ultraviolet (and into X-rays).  In this paper, we will make use of the $UBVRIJHK_S$ bands.  Furthermore, NGC 188 has precise (ground-based) proper motions \citep{platais} and radial velocities \citep{gellerrvs}, allowing for a clean separation of the field and cluster stars, with minimal contamination \citep[e.g. see][]{gellerrvs}.  This sample of cluster members was used by \citet{shane}, using a similar Bayesian method as we employ here, to study the cluster single stars and derive cluster parameters. 

Furthermore, NGC 188 has a comprehensive and complete radial velocity survey of the solar-type binary stars \citep{gellerorbits, gellerhardbin}, including nearly 100 binary orbital solutions.  The Geller et al.\ papers study the binary frequency and distributions of orbital parameters (including the mass-ratio distribution).  For the double-lined binaries, these orbital solutions provide direct measurements of the respective mass ratios.  These detailed spectroscopic observations are essential to verify our photometric analysis methods.    

Unlike in \citet{shane}, here we include the binary stars in our Bayesian analysis of the cluster, with the goals of (a) comparing the results of our analysis with those of \citet{gellerorbits} and \citet{gellerhardbin}, and (b) uncovering additional binaries (e.g., at larger orbital periods)  
that were not accessible to the Geller et al.\ study. 

The remainder of this study is organized as follows: In Sect.~\ref{datasect}, we describe our observational catalog and the cuts used to select main sequence binaries.  In Sect.~\ref{bayessect}, we describe the Bayesian methodology used to produce posterior distribution functions (PDFs) of the primary and putatuve secondary masses for cluster members, and in Sect.~\ref{simsect}, simulations are used to characterize biases and uncertainties in our analysis technique when recovering properties of the cluster binary population.  In Sect.~\ref{resultsect}, we quantify several properties of the NGC 188 main sequence binary population and compare them to existing measurements, and our results are summarized in the final section.  

\section{Observational Data \label{datasect}}
\subsection{Sample Selection \label{cutsect}}

\begin{figure}
\plotone{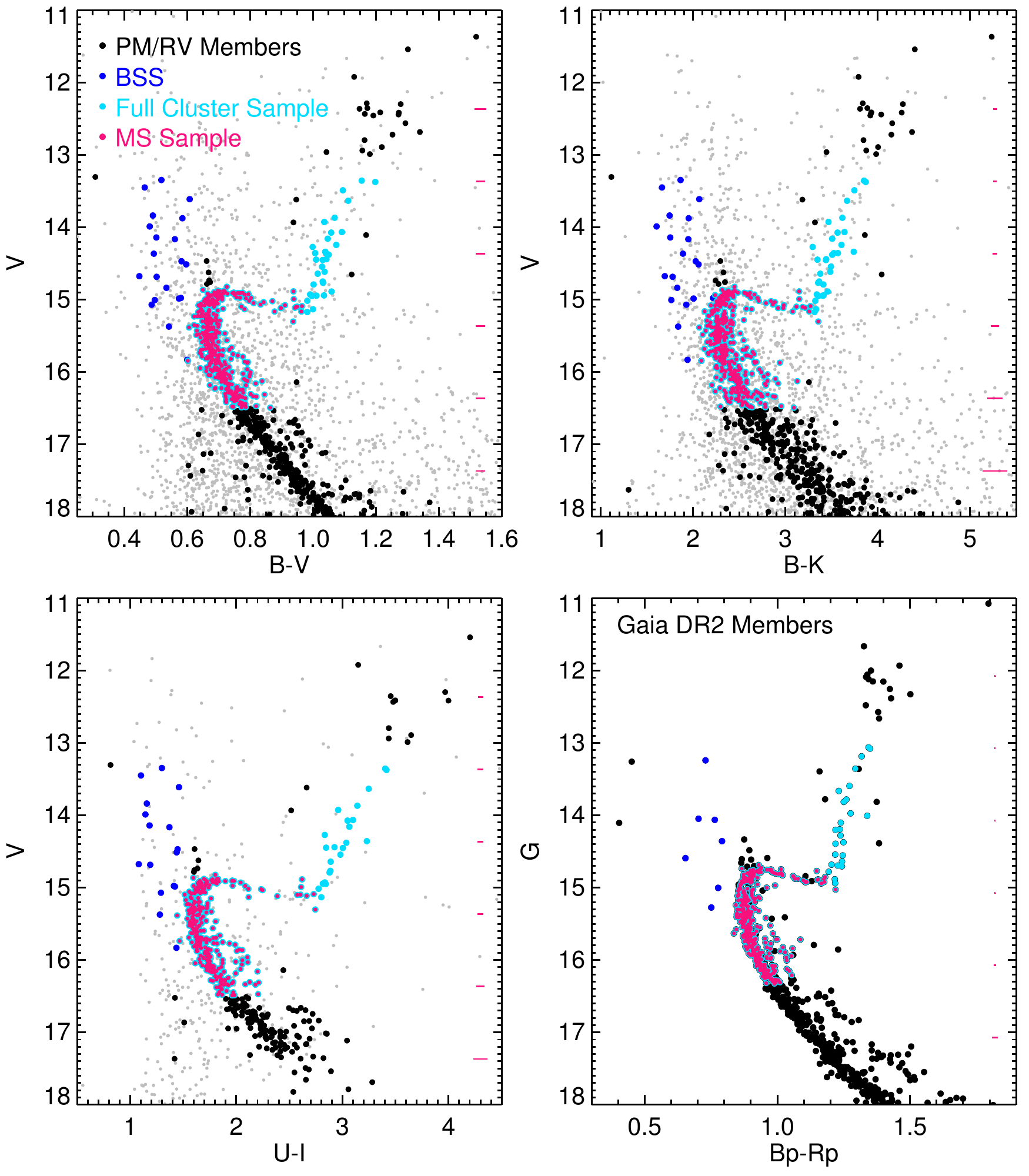}
\caption{Color-magnitude diagrams in four color-magnitude planes, illustrating the full photometric catalog ($UBVRI$ from \citealt{stetson} and $JHK_{S}$ from 2MASS).  Stars in grey fail our 50\% membership probability cut and are excluded from our analysis, and the remainder (black, blue, magenta and cyan) pass this cut.  Blue circles are BSS from \citet{gosnell15} which are also excluded from our analysis.  Cyan filled circles are included in our full cluster sample (used to derive cluster and stellar parameters), while only stars also in magenta are included in our analysis of the main sequence binary population.  In the lower right panel, only members from \citet{gaiacat} are shown.  Median photometric errors in magnitude bins are shown on the right-hand side of each CMD. \label{cmdobsfig}}
\end{figure}

Our photometric catalog consists of $UBVRI$ photometry from \citet{stetson}, supplemented with $JHK_{S}$ photometry from the Two Micron All Sky Survey (2MASS) Point Source Catalog \citep{skrutskie}.  Prior membership probabilities were taken from the proper motion study of \citet{platais} and the radial velocity studies of \citet{gellerrvs,gellerorbits} and \citet{gellerhardbin}.  When both proper motion and radial velocity membership probabilities were available, we took a simple arithmetic mean of the two values to derive the final membership probability, $P_{mem}$, which was used as a prior value on the membership probability of each star.  With the multi-band catalogs and a single value of $P_{mem}$ available for each star, we then further restricted the sample employed in our Bayesian analysis using the following cuts:

\begin{enumerate}

 \item All known blue straggler stars (BSS) from \citet[][and references therein]{gosnell15} were eliminated, as well as additional candidates based on their color-magnitude loci, rejecting stars brightward and blueward of the main sequence turnoff with $(B-V)$$<$0.6 and $V$$<$14.8.  Because our Bayesian technique functions by comparing observed magnitudes in each bandpass to their expected values based on single-age evolutionary models, the inclusion of such non-canonical evolutionary products would affect the posterior distributions of cluster properties.  Since we are concerned with main sequence binaries here, we need not try to fit these systems.

 \item We include only the stars with 13$\leq$V$\leq$16.5, 
 which effectively serves as a S/N cut in the photometric catalog while preserving the cluster main sequence down to the magnitude limit of the available radial velocity data.  This full sample of 432 stars includes red giant branch (RGB) stars, shown in cyan in Fig.~\ref{cmdobsfig}, which are crucial to the simultaneous determination of cluster parameters (distance, reddening, metallicity and age), although we further restrict our analysis of the main sequence binary population in Sects.~\ref{simsect} and \ref{resultsect} to the 388 stars with $(B-V)$ $<$ 0.98, corresponding to primary masses 0.95$\leq M \leq$1.15$M_{\sun}$, shown in magenta in Fig.~\ref{cmdobsfig}.  

 \item We only include stars with $P_{mem}\geq$50\%.

 \item Each star must have photometry in at least the $B$ and $V$ filters.  

\end{enumerate}

A histogram of membership probabilities (from radial velocities \textit{and} proper motions) is shown in the left panel of Fig.~\ref{gaiapmfig}, with all stars present in the full $UBVRIJHK_S$ catalog in black and the stars passing our CMD cuts in blue.  As most stars are quite confidently identified as members or non-members, the exact location of our intermediate cut in membership probability has
little effect on the resulting sample.

\subsection{Member Selection: Comparison to Gaia DR2}

\begin{figure}
\gridline{\fig{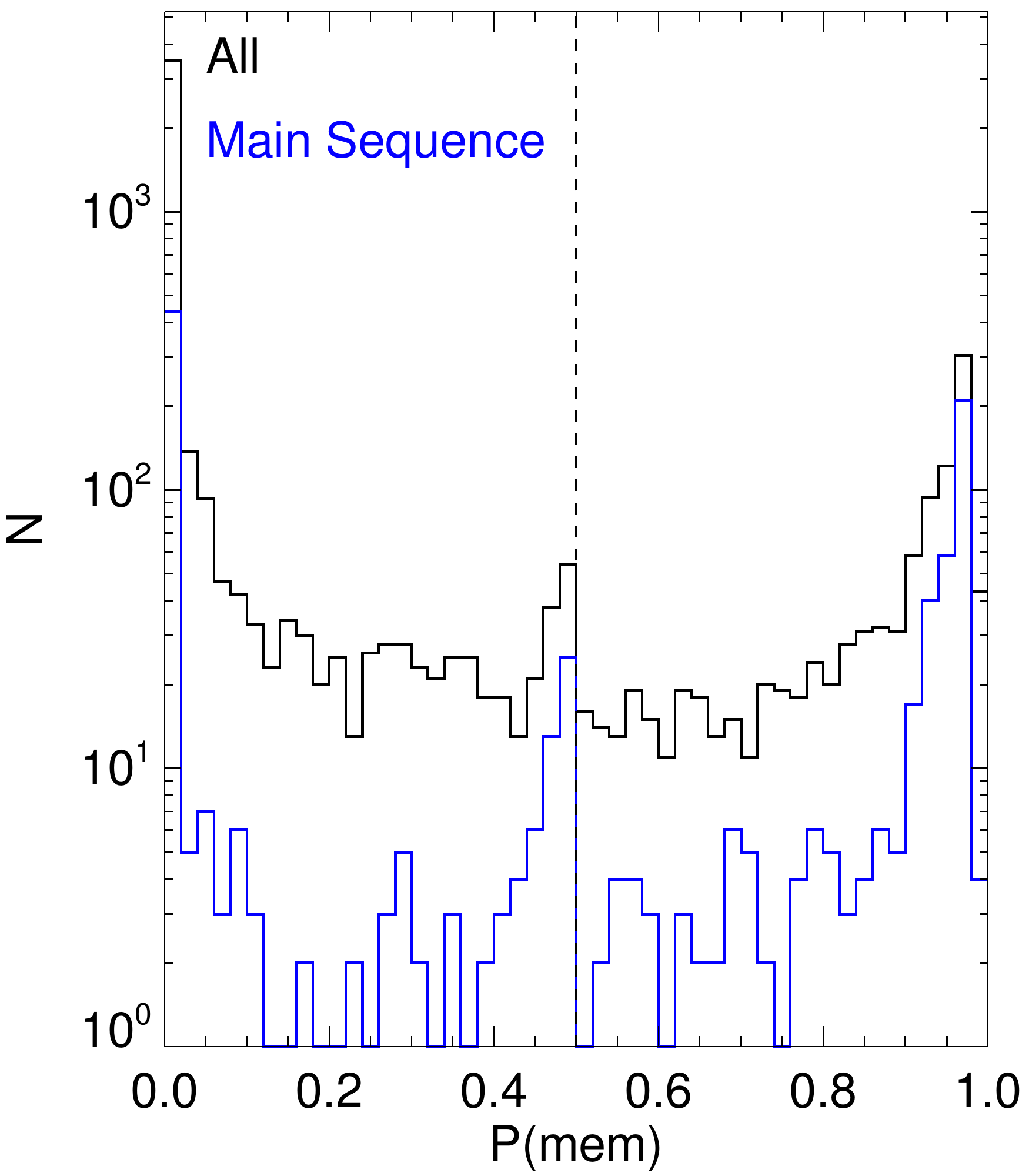}{0.3\textwidth}{}
		  \fig{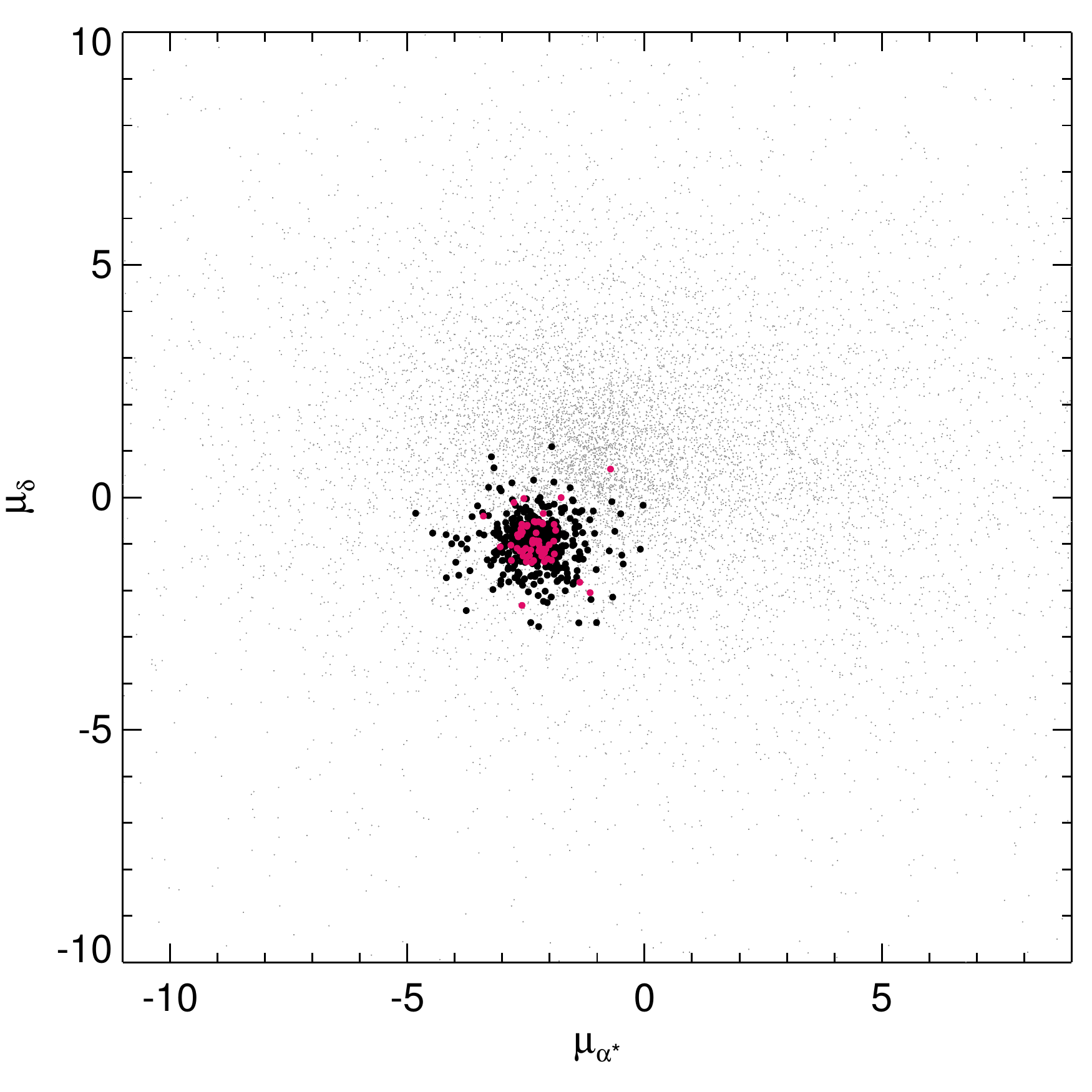}{0.35\textwidth}{}
		  \fig{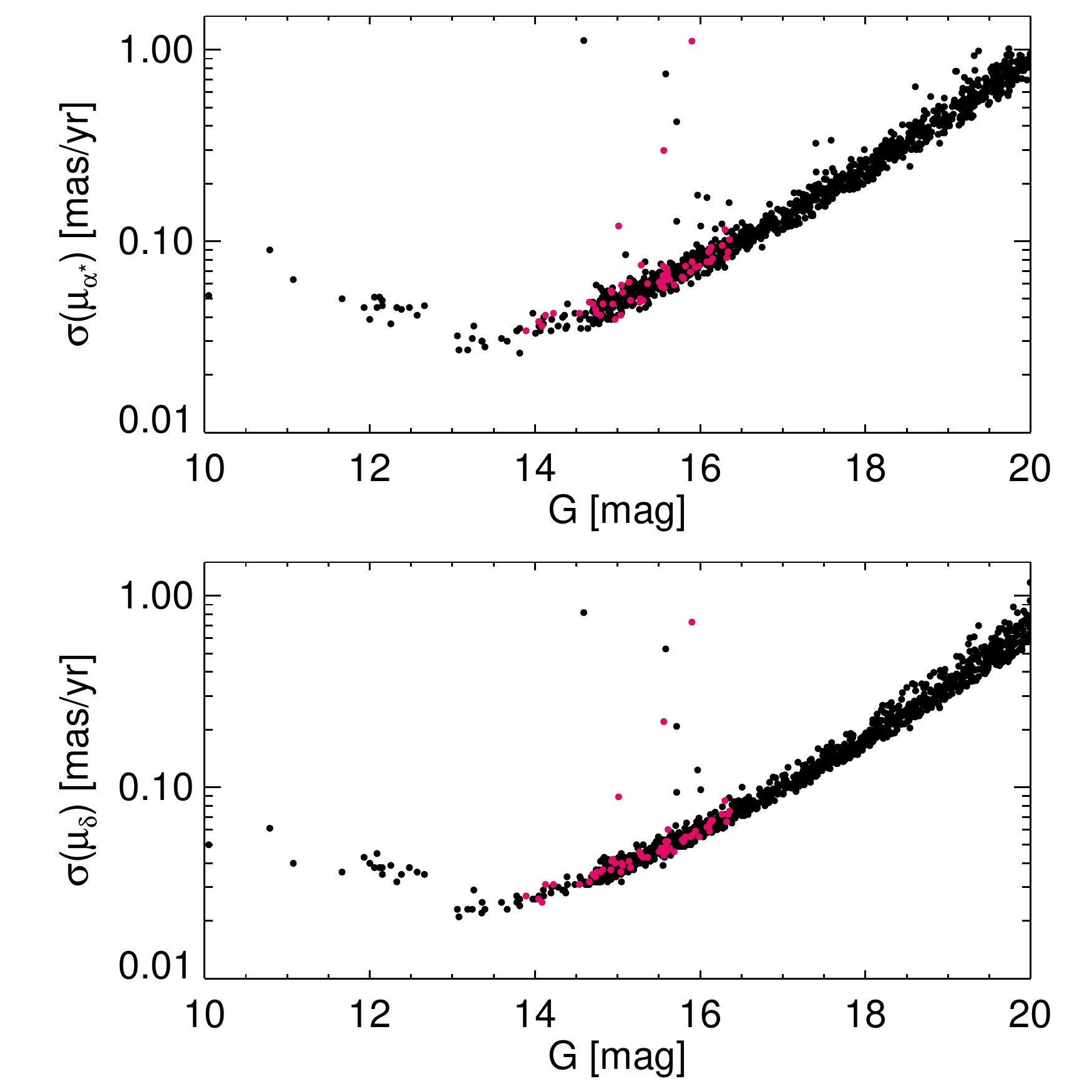}{0.35\textwidth}{}}
\caption{\textbf{Left:} Histograms of membership probabilities from ground-based proper motions and radial velocities for all stars in our photometric catalog (black) and main sequence stars (blue) passing the CMD cuts described in Section~\ref{cutsect}.  Our 50\% cut in membership probability for inclusion in the final sample is shown as a vertical dashed line. \textbf{Middle:} Vector point diagram illustrating the proper motions of all stars in \textit{Gaia} DR2 within 36.2$\arcmin$ (the largest distance from the cluster to which they detect members) of NGC 188 (grey points), \textit{Gaia} members in black, and the stars in our catalog which passed all of our photometric and membership cuts but are not present in the \textit{Gaia} list of cluster members in magenta.  \textbf{Right:} Proper motion uncertainties as a function of $G$ magnitude.  Symbols are as in the middle panel.  Our proper-motion and/or radial-velocity members (magenta) missing from the \textit{Gaia} selection of cluster members (black) have proper motions and uncertainties consistent with membership, and were rejected from the \textit{Gaia} DR2 cluster sample due to failing the quality cuts described in \citet{gaiacat}. \label{gaiapmfig}}
\end{figure}

The membership information provided by the second data release of the \textit{Gaia} satellite provides an opportunity to validate our membership criteria.  Of the 432 stars in our final sample, 61 (14\%) were not identified as members in the cluster catalog provided by \citet{gaiacat}.  However, all of these 61 stars are present in \textit{Gaia} DR2, and upon closer inspection, they were not listed as members due to the quality cuts described in Section~2.1 and Appendix B of \citet{gaiacat}.  Meanwhile, in Fig.~\ref{gaiapmfig} we illustrate that the proper motions and proper motion errors of these stars, shown in magenta compared to \textit{Gaia} members in black, give no reason to suspect that any significant fraction of them are non-members.  We further checked the remaining three parameters and uncertainties (RA, Dec, parallax) from the five-parameter astrometric solution and similarly found that the error distributions of the stars excluded from the \textit{Gaia} membership list did not deviate discernibly from that of the \textit{Gaia}-selected members.

\section{Bayesian Determination of Cluster and Stellar Parameters \label{bayessect}}

We employ BASE-8, a tool for fitting and characterizing observations of star clusters and even individual stars using multi-band photometry \citep{vonhippel, degennaro, vandyk, stein, stenning}. BASE-8 functions by comparing observed photometric catalogs to a user-supplied grid of isochrones, with the goal of providing posterior distribution functions (PDFs) of both cluster parameters and stellar parameters.  A crucial advantage of BASE-8 is that it can handle input and output values on a star-by-star basis rather than simply providing cluster parameters which best reproduce a given photometric catalog.  Specifically, in addition to the isochrone grid, cluster-wide inputs include Gaussian priors on distance $(m-M)_{V}$, reddening $A_{V}$, metallicity [Fe/H], and a flat prior on Log age.  In addition, the input photometric catalog to be compared with the isochrone grid can contain missing or incomplete photometry for any subset of stars in any combination of the selected filters, and can also contain individual membership probabilities on a star-by-star basis.  Given these inputs, a Bayesian maximum likelihood analysis is performed which yields PDFs for the four cluster parameters distance, reddening, metallicity and age, \textit{and} star-by-star information (described below).  

BASE-8 also has the advantage of (optionally) exploiting user-supplied prior information on the cluster parameters, and so for each of the four cluster parameters (distance, reddening, metallicity and age) one can make use of both a prior distribution and a constraint on that value, given as 
an input Gaussian mean and standard deviation (in logarithmic space).  For NGC 188, we assume the same input prior means and uncertainties as \citet{shane} based on main-sequence fitting to observed CMDs by \citet{ata}, which are given in Table \ref{priortab}\footnote{A flat prior distribution is assumed for Log age over the DSED model limits, i.e. from 1 Gyr to 15 Gyr.  The uninformative age prior means that the likelihood (the fitting of DSED isochrones to the multi-band photometry) dominates the age posterior, and sensitivity tests over a range of priors by \citet{shane} found that reasonable variations for these priors yielded negligible differences in the resulting posterior distributions.}.
For our isochrone grid, we use solar-scaled models from the Dartmouth Stellar Evolution Database \citep[DSED;][]{dsed}.

\begin{deluxetable}{lcc}
\tablecaption{Prior Means and Standard Deviations for Cluster Parameters \label{priortab}}
\tablehead{
\colhead{Parameter} & \colhead{Value} & \colhead{$\sigma$}}
\startdata
$(m-M)_{V}$ & 11.44 & 0.3 \\
$A_{V}$ & 0.3 & 0.1 \\
$[Fe/H]$ & -0.03 & 0.3 \\
Log Age\tablenotemark{a} & & inf \\
\enddata
\tablecomments{Prior values are the same as those employed by \citet{shane}, based on main sequence fitting by \citet{ata}.}
\end{deluxetable}

\subsection{Main Sequence Binary Mass Ratios}

Importantly, within the posterior distribution, the resulting star-by-star information is also available, yielding PDFs of stellar mass for each star and its companion, over all MCMC iterations.  No \textit{a priori} assumption is made regarding the binary fraction of the cluster.  Moreover, in each MCMC iteration a given star is not assumed \textit{a priori} to be single, so that constraints on binarity for any star come from the full PDF of primary and putative secondary masses over all MCMC iterations.  In each MCMC iteration, all available multi-band photometry for each star is compared to the model grid to determine the mass of both components. 

In other words, BASE-8 is broadly equivalent to SED fitting on a star-by-star basis, yet BASE-8 simultaneously performs a hierarchical fit where all cluster stars have the same cluster-wide properties (age, metallicity, etc.) at a given iteration.  The primary and secondary stellar mass PDFs therefore not only incorporate uncertainties in the fits of the single or binary pair to the photometry, but also incorporate the uncertainties in the cluster parameters and benefit from all stars being fit simultaneously.

A star that is extremely unlikely to have a companion would have a PDF of secondary masses $M_{2}$ centered very close to zero with little scatter.  Therefore, to characterize the mass ratios of main sequence members of NGC 188, we use the mass ratio $q = \frac{M_{2}}{M_{1}}$, and use the PDF of the secondary mass $M_{2}$ as a criterion to identify likely binaries.  Specifically, using the distribution of $M_{2}$ over all MCMC iterations, we define candidate binaries identified by BASE-8 as those stars for which $M_{2} > 3 \sigma\left(M_{2}\right)$, where $M_{2}$ is the median value taken over all iterations, and $\sigma\left(M_{2}\right)$ is the 16th-84th fractile of the PDF of $M_{2}$ over all iterations.  

This criterion, and any resulting biases in either the binary fraction $f(bin)$ and/or input vs. recovered values of $q$, 
can be validated using simulations.

\section{Validation Through Simulations \label{simsect}}
\subsection{Input Parameters \label{siminputsect}}
\begin{figure}
\gridline{\fig{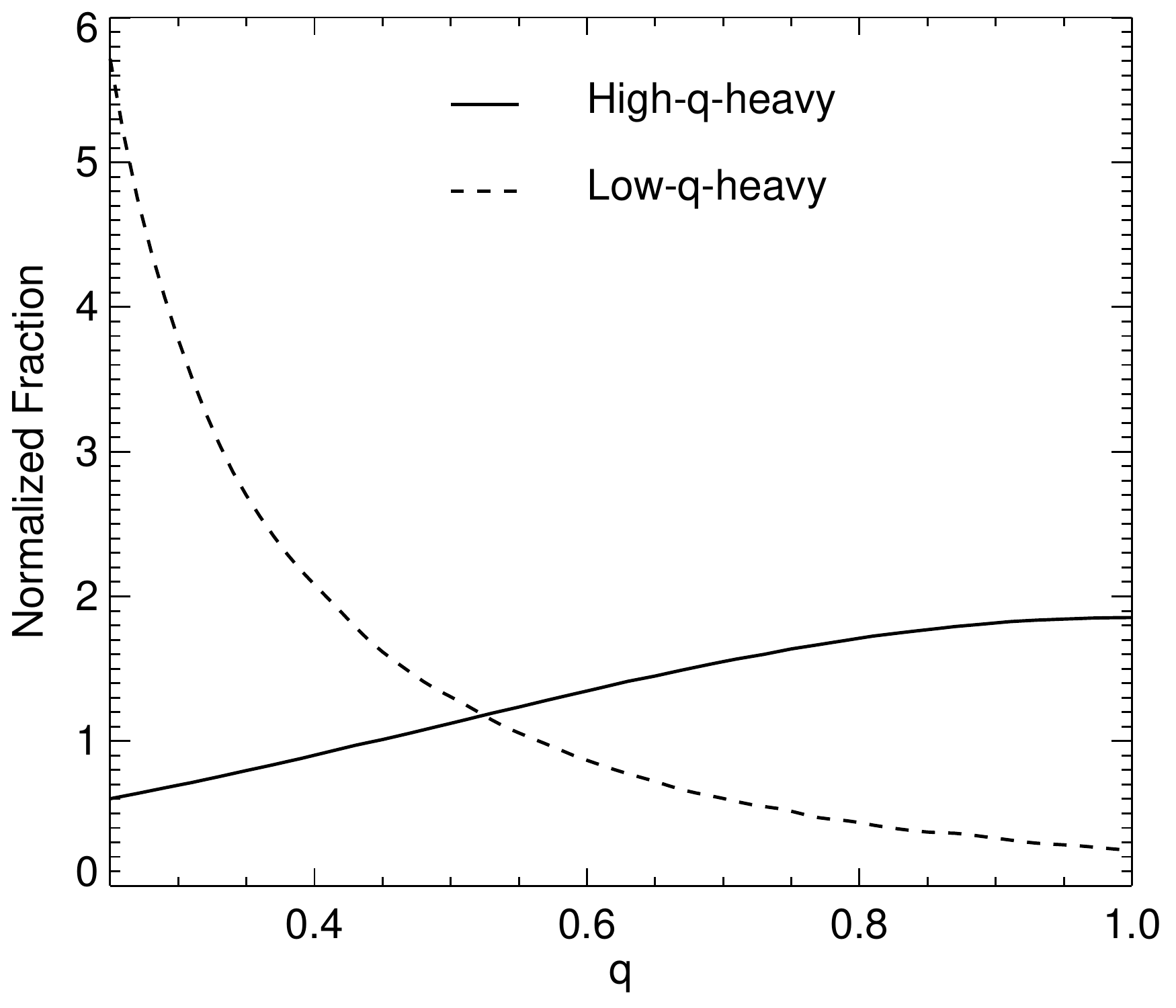}{0.32\textwidth}{}
		  \fig{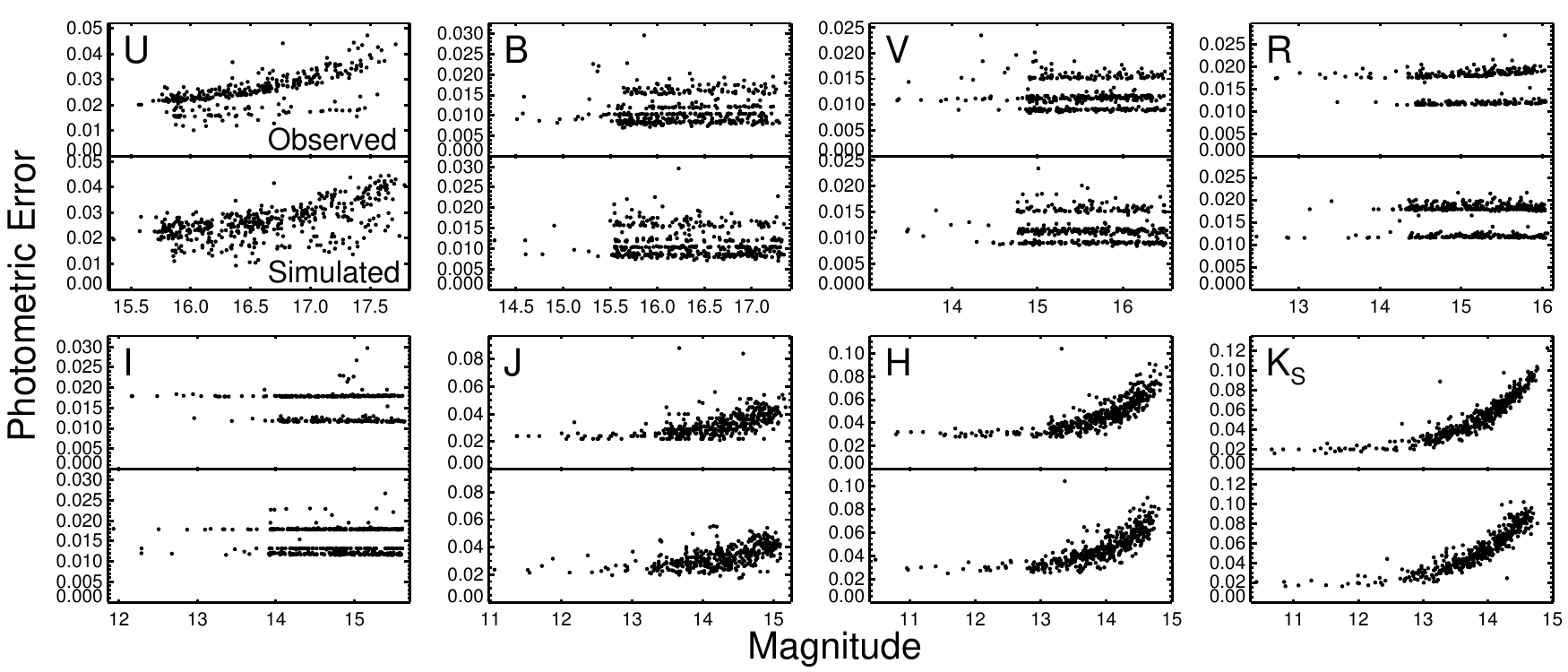}{0.65\textwidth}{}}
\caption{\textbf{Left:} The two mass ratio distributions used for our simulations.  For each mass ratio distribution, multiple cluster realizations are generated with three different input binary fractions and run through BASE-8 to assess our ability to recover main sequence binary properties (see text for details).  \textbf{Right:} For each of the eight filters in our observed catalog, we show the observed photometric errors as a function of magnitude (upper panel) and the simulated photometric errors (lower panel) which were applied to a realization of a simulated cluster.
 \label{simphoterrfig}}
\end{figure}

Before running BASE-8 on the true NGC 188 data, we generate multiple artificial star cluster realizations, tailored to mimic our observational catalog as closely as possible.  By generating artificial clusters in which the parameters of the main sequence binary population are known, we may assess our ability to recover the binary fraction $f(bin)$ and the distribution of main sequence binary mass ratios $q$.  Artificial clusters are generated using the following inputs: 

\begin{itemize}

    \item{\textbf{Cluster Parameters:} We use an identical set of DSED isochrones as those assumed to model the real data.  These isochrones are shifted to the observational plane using the same values of distance, reddening, $[Fe/H]$ and age assumed as the priors given in Table \ref{priortab}.  
    An artificial star cluster is then generated by distributing stars over the isochrone with a \citet{salpeter} initial mass function (IMF), although our results do not depend sensitively on this choice given the 
    mass range (see Sect.~\ref{cutsect}) covered by our sample.}   
    
    \item{\textbf{Binary Fraction:} An input binary fraction is used to specify the number of binary stars.  We generate artificial clusters with input binary fractions of $f(bin)$=20, 50 and 80\%, and generate multiple realizations for each of these values of $f(bin)$ to improve Poisson statistics.} 
    
    \item{\textbf{Mass Ratio Distribution:} For a given input $f(bin)$, we generate multiple cluster realizations for each of two mass ratio \textit{distributions}, which we designate as \textquotedblleft low-q-heavy\textquotedblright{}, drawn from a \citet{ktg} initial mass function, and \textquotedblleft high-q-heavy\textquotedblright{}, drawn from a Gaussian distribution of mass ratios with a mean of one and a standard deviation of 0.5.  
    These mass ratio distributions are shown in the left-hand panel of Fig.~\ref{simphoterrfig}.  As the true mass ratio distributions of main sequence binaries in open clusters remain fairly poorly constrained \citep[e.g.][]{duchene,moe}, our intent is not to use the simulations to constrain the detailed nature of the mass ratio distribution in NGC 188; rather, the use of two schematically distinct mass ratio distributions allows us to assess whether such a difference impacts our results regarding the properties of NGC 188 main sequence binaries (discussed later in Sects.~\ref{binfracsect} and \ref{massratsect}).}  
    
    \item{\textbf{Photometric Errors and Cuts:} To permit a direct comparison to the observational results, stars in an artificial cluster realization are offset using Gaussian deviates of their photometric errors as a function of magnitude, drawn from the observed error distribution for each filter, plus an additional random scatter $\sim$0.002 mag which we found further improved the similarity between the observed and artificial photometric error distributions.  A direct comparison between the photometric errors applied to a simulated cluster versus those in the observational catalog is shown for all 8 filters in Fig.~\ref{simphoterrfig}.  Lastly, CMD cuts are applied in an identical fashion as to the observational catalog as listed in Sect.~\ref{cutsect}, and the total number of stars passing these cuts is constrained to be identical to our NGC 188 observational catalog.  However, for each artificial cluster realization, we retain the input parameters (i.e.~primary mass and mass ratio) for stars eliminated from the output sample by our CMD cuts, as these cuts may bias the observed binary fraction and mass ratio distribution compared to a primary-mass-limited sample (discussed below in Sect.~\ref{biassect}).}
\end{itemize}    

In Fig.~\ref{simcmdoutfig} we show CMDs of three example simulated cluster realizations, varying both the input binary fraction (left and center panels) and mass ratio distribution (center and right panels).  Input binaries are color-coded by their input mass ratio $q(in)$ according to the colorbar at the top of the plot, and binaries which were successfully recovered by BASE-8 are marked with diamonds.  While it is qualitatively apparent that the majority of input binaries are recovered successfully, particularly those with high mass ratios, we examine the simulation results quantitatively with the goal of assessing how well we are able to recover the main sequence binary fraction, and how this depends on the input parameters (i.e.~binary fraction and mass ratio distribution).  However, in order to do so, we must account for selection effects that bias our recovered binary properties due, in part, to the use of a CMD-selected main sequence sample.  

\subsection{Accounting for Photometric Selection Biases \label{biassect}}

There are several 
selection effects that may influence recovered main sequence binary properties:

\begin{enumerate}
    
    \item Due to the verticality of the main sequence in the vicinity of the main sequence turnoff, high mass ratio binaries whose single star analogs lie slightly faintward of the turnoff are preferentially \textquotedblleft hidden\textquotedblright{} in the turnoff, with CMD loci coincident with single turnoff stars.  For example, by concatenating all simulation runs together, we find that  $>$70\% of the unrecovered input binaries with $q(in) >$0.5 have $V <$15.1.  However, the reason why this effect is not immediately apparent from any individual simulation run (i.e.~in Fig.~\ref{simcmdoutfig}) is that the \textit{fraction} of $q(in) >$0.5 input binaries which are unrecovered remains relatively small, 26$\pm$7\% in the vicinity of the main sequence turnoff ($V <$15.1) and $<$10\% over the remainder of the main sequence. 

    \item The use of a bright magnitude limit to exclude blue stragglers from our main sequence sample also excludes high-mass-ratio binaries whose single-star analogs (i.e.~with the same primary mass $M_{1}$) lie at or slightly below the main sequence turnoff. 
    
    \item The bias caused by imposing a \textit{faint} magnitude limit is due to the inclusion of high-mass-ratio binaries whose single-star analogs are absent because they lie faintward of the faint magnitude cutoff.   
    
\end{enumerate}

An unbiased analysis of the cluster main sequence binary properties should be performed using a primary-mass-limited sample, but we are faced with the problem that the (true) primary mass is unknown in our observed NGC 188 catalog.  Therefore, to include the effects of the aforementioned three biases in our analysis of the simulations, we proceed as follows: We designate an \textit{input} sample which is defined by a lower limit on primary mass.  Specifically, we require $M \geq$ 0.95$M_{\sun}$, corresponding to a single star magnitude of V=16.5, down to which the CMD-selected sample includes stars of all mass ratios, including singles (modulo detection incompleteness and photometric errors, both included in the simulations).  The output sample, on the other hand, is necessarily based on CMD cuts because we do not have access to the true primary masses of stars in the observed catalog.  In this way, comparison of simulation input binary properties (based on primary mass) versus recovered binary properties (based on CMD cuts) incorporates the effects of the various biases described above.   

\subsection{Recovery of Binary Fraction \label{simbinfracsect}}

\begin{figure}
\gridline{\fig{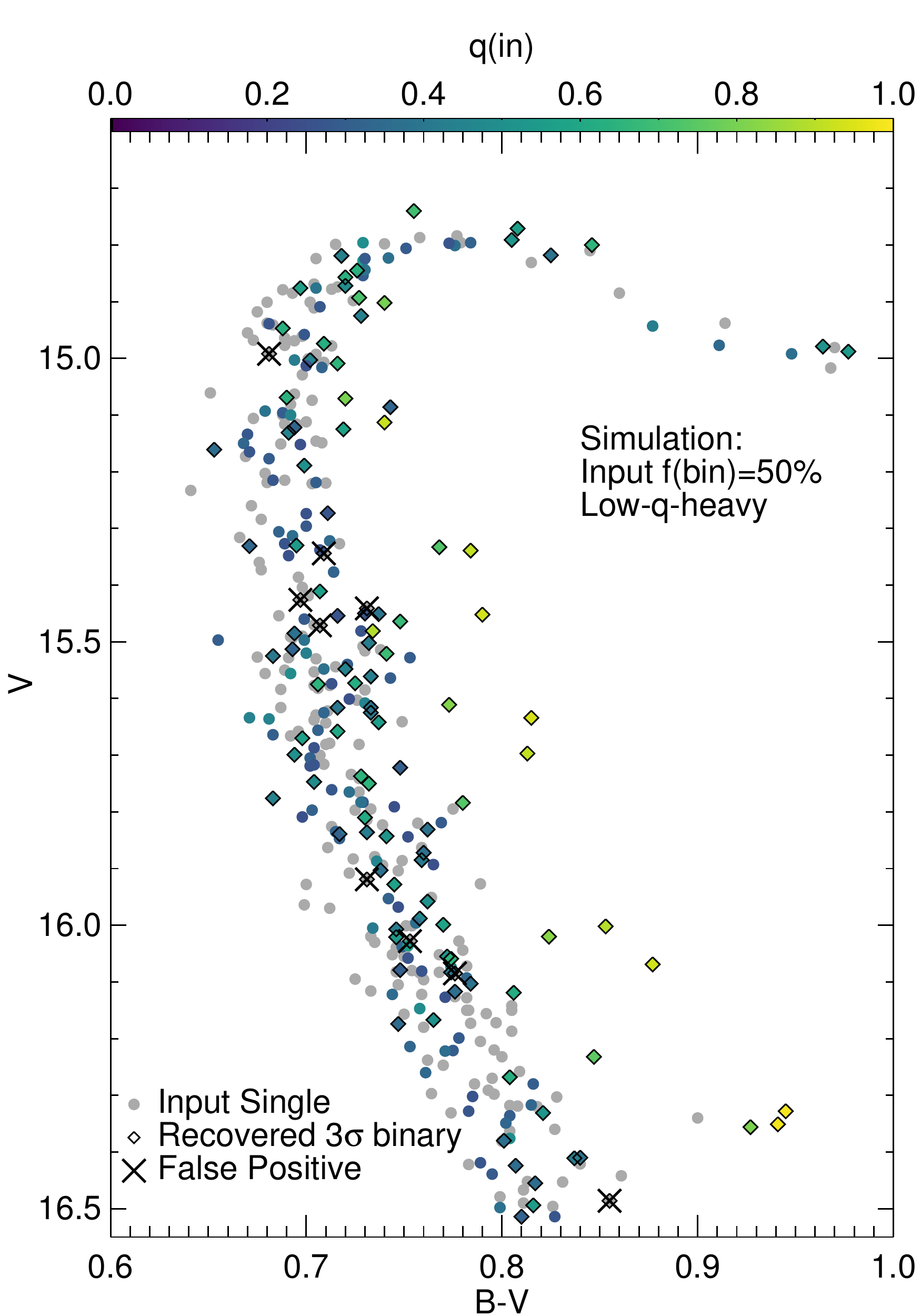}{0.33\textwidth}{}
		  \fig{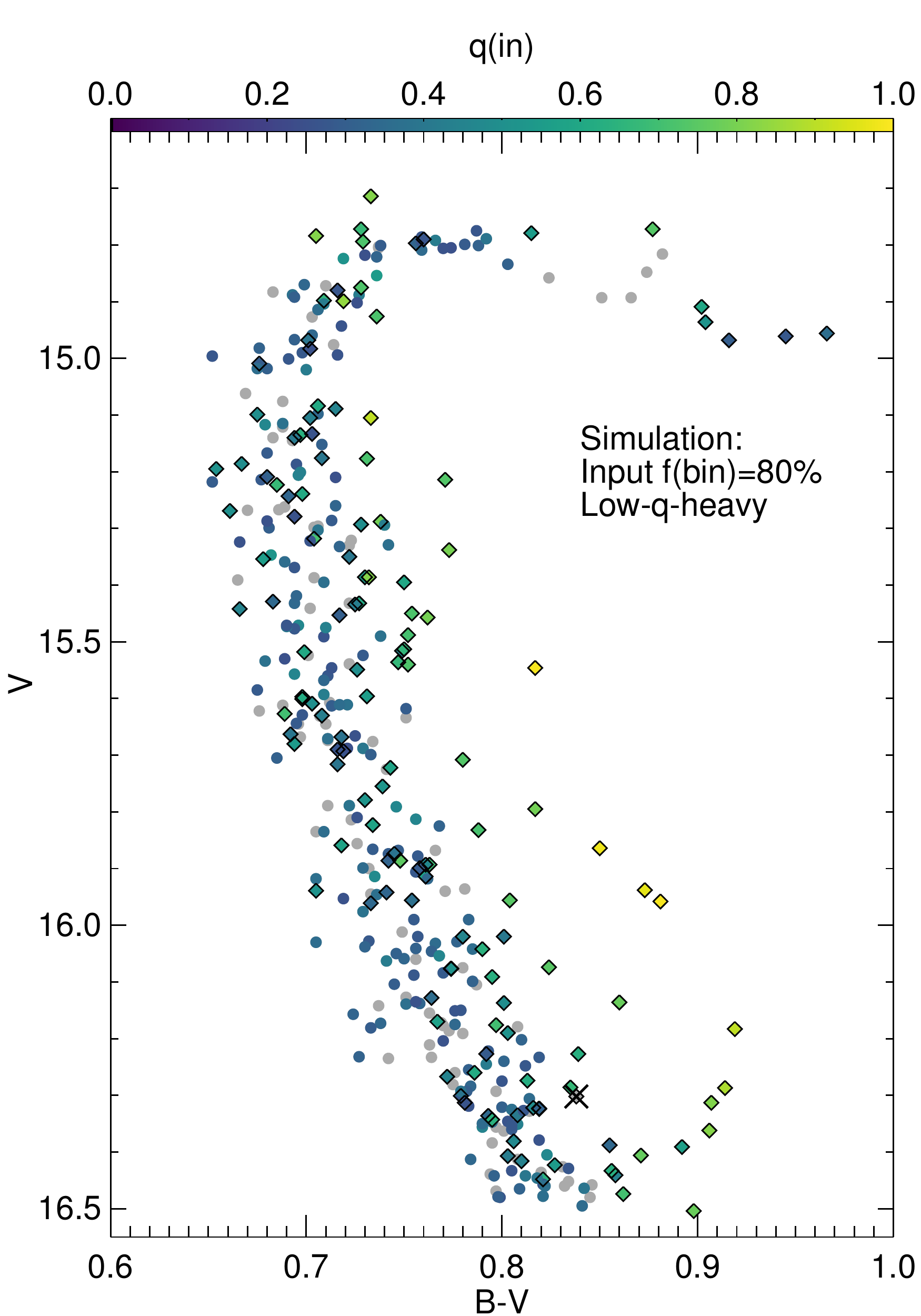}{0.33\textwidth}{}
		  \fig{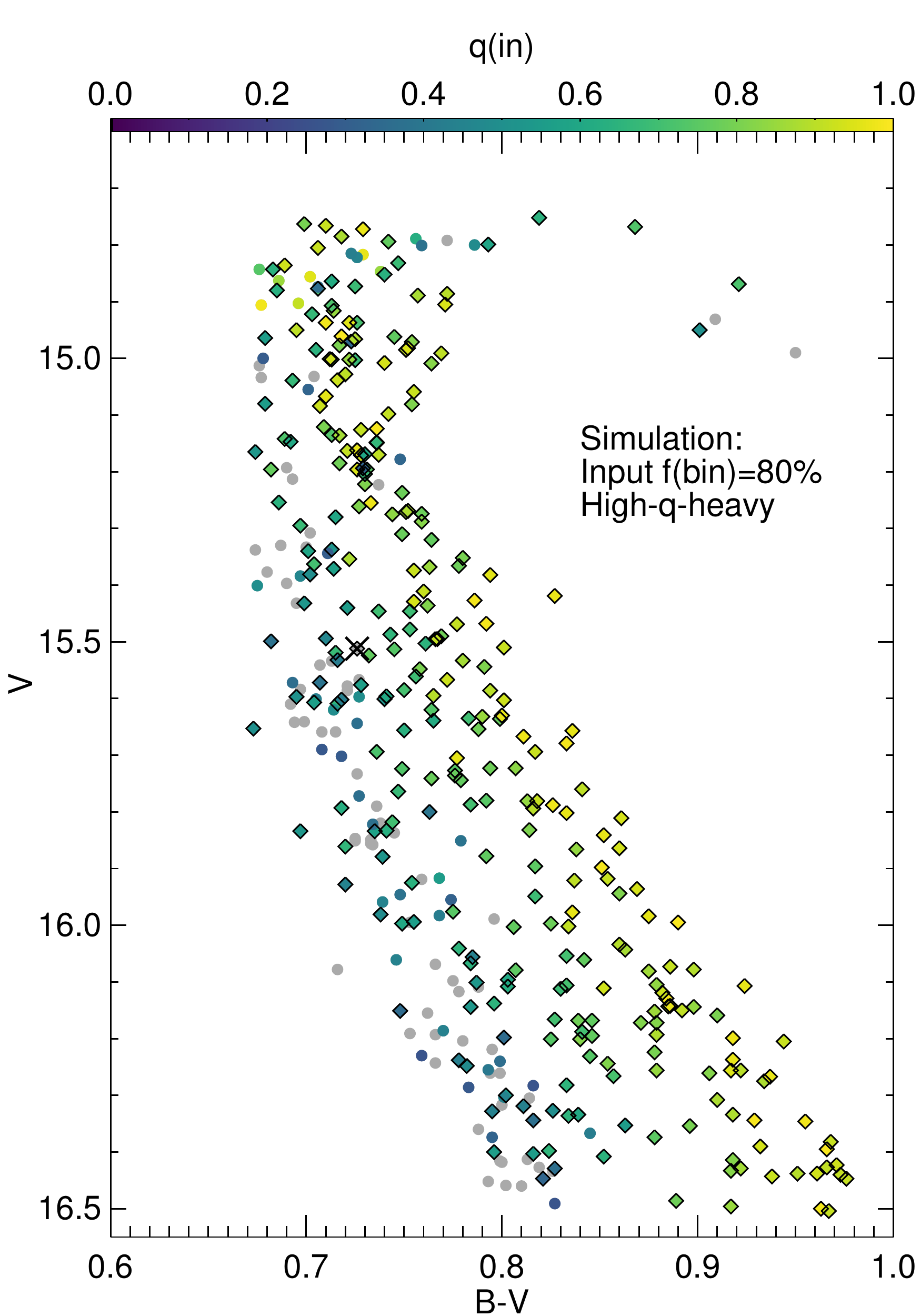}{0.33\textwidth}{}}
\caption{Example CMDs of simulated clusters in which stars are color-coded by their input mass ratio $q(in)$.  Single stars are shown in grey, binaries successfully recovered by BASE-8 are marked with diamonds, and input singles misidentified as binaries are marked with crosses.  The left and center panels compare two example cluster realizations, both with a low-q-heavy input mass ratio distribution but differing binary fractions of 50\% (left) and 80\% (center), while the center and right-hand panels compare two clusters with input binary fractions of 80\% but different input mass ratio distributions.
 \label{simcmdoutfig}}
\end{figure}

In Table \ref{simresultstab}, we summarize the results of the simulations, broken down by input mass ratio distribution and by input binary fraction $f(bin,in)$, given in the first two columns.  The next column gives the recovered binary fraction $f(bin,out)$, which is simply the ratio of binaries passing our $M_{2} >$3$\sigma(M_{2})$ criterion compared to the size of the CMD-selected main sequence sample.  The quoted uncertainties are the quadrature sum of Poissonian uncertainties and run-to-run variations among multiple cluster realizations with identical input binary population properties (mass ratio distribution and binary fraction).  Importantly, the input binary fraction $f(bin,in)$ refers to the input primary-mass-limited sample, so that a direct comparison between $f(bin,in)$ and $f(bin,out)$ includes not only the selection effects described in Sect.~\ref{biassect}, but also any systematics due to detection incompleteness, as well as the presence of false positives (shown as large crosses in Fig.~\ref{simcmdoutfig}), which are input single stars that are recovered as binaries, meaning that they satisfy our output $M_{2} >$3$\sigma(M_{2})$ criterion.  To clearly assess how the relationship between $f(bin,in)$ and $f(bin,out)$ may depend on the assumed binary fraction or mass ratio distribution, we also give the quantity $f(rec)$, which is simply the ratio $f(bin,out) / f(bin,in)$.  

It is clear from the $f(rec)$ values that binaries with high mass ratios are recovered more successfully, as the fraction of recovered binaries drops by nearly half when they are assigned input mass ratios drawn from a low-q-heavy distribution.  With this in mind, the remaining columns of Table \ref{simresultstab} give results restricted to binaries with 
mass ratios $q >$0.5, such that $f(bin,in,q > 0.5)$ refers to the fraction of input binaries with input mass ratios $q>$0.5 in the primary-mass-selected sample, compared to $f(bin,out,q > 0.5)$, which is the fraction of binaries in the CMD-selected sample which are recovered successfully \textit{and} have \textit{output} mass ratios $q >$0.5.  A comparison between $f(rec)$ over all mass ratios against $f(rec,q > 0.5)$ immediately reveals at least two advantages to restricting our analysis to a $q >$0.5 sample: First, the binary fraction is recovered much more accurately, 
to within Poissonian uncertainties in over 90\% of individual simulation runs, while the full ($q>0$) binary fraction was recovered within Poissonian uncertainties in less than 10\% of the runs.  Furthermore, the difference between output and input $q >$0.5 binary fraction had a mean of 0.009$\pm$0.007 (0.000$\pm$0.011) and a median absolute deviation (MAD) of 0.009 (0.020) for the low-q-heavy (high-q-heavy) mass ratio distributions.  
A second, related advantage to a $q >$0.5 sample is that $f(rec)$ shows a drastically reduced dependence on the input parameters of the binary population.  For example, Table \ref{simresultstab} reveals that with no restrictions on $q$, $f(rec)$ is significantly correlated with both the input binary fraction $f(bin,in)$ and also the mass ratio distribution.  Conversely, for the $q >$0.5 sample, the correlation with $f(bin,in)$ is no longer statistically significant, and the dependence on mass ratio distribution is marginally ($\sim$1$\sigma$) significant.   

\begin{deluxetable}{lcccccc}
\tablecaption{Simulation Results \label{simresultstab}}
\tablehead{
\colhead{$q$ Distribution} & \colhead{$f(bin,in)$} & \colhead{$f(bin,out)$} & \colhead{$f(rec)$} & \colhead{$f(bin,in,q$$>$$0.5)$} & \colhead{$f(bin,out,q$$>$$0.5)$} & \colhead{$f(rec,q$$>$$0.5)$}}
\startdata
  Low-q-heavy &      0.8 & 0.353$\pm$0.018 & 0.442$\pm$0.019 & 0.202   & 0.214$\pm$0.022 & 1.059$\pm$0.050 \\
  Low-q-heavy &      0.5 & 0.271$\pm$0.021 & 0.535$\pm$0.025 & 0.157   & 0.169$\pm$0.024 & 1.076$\pm$0.070 \\
  Low-q-heavy &      0.2 & 0.157$\pm$0.015 & 0.708$\pm$0.035 & 0.059   & 0.070$\pm$0.016 & 1.198$\pm$0.247 \\
  Low-q-heavy & Combined & 0.312$\pm$0.065 & 0.470$\pm$0.091 & 0.177   & 0.188$\pm$0.050 & 1.061$\pm$0.051 \\
 High-q-heavy &      0.8 & 0.698$\pm$0.043 & 0.855$\pm$0.042 & 0.669   & 0.650$\pm$0.034 & 0.972$\pm$0.039 \\
 High-q-heavy &      0.5 & 0.468$\pm$0.029 & 0.914$\pm$0.034 & 0.437   & 0.423$\pm$0.024 & 0.968$\pm$0.060 \\
 High-q-heavy &      0.2 & 0.263$\pm$0.018 & 1.190$\pm$0.108 & 0.182   & 0.193$\pm$0.015 & 1.060$\pm$0.067 \\
 High-q-heavy & Combined & 0.499$\pm$0.193 & 0.918$\pm$0.164 & 0.451   & 0.446$\pm$0.201 & 0.988$\pm$0.062 \\
\enddata
\end{deluxetable}

\subsection{Recovery of Mass Ratios \label{simmassratsect}}
\begin{figure}
\gridline{\fig{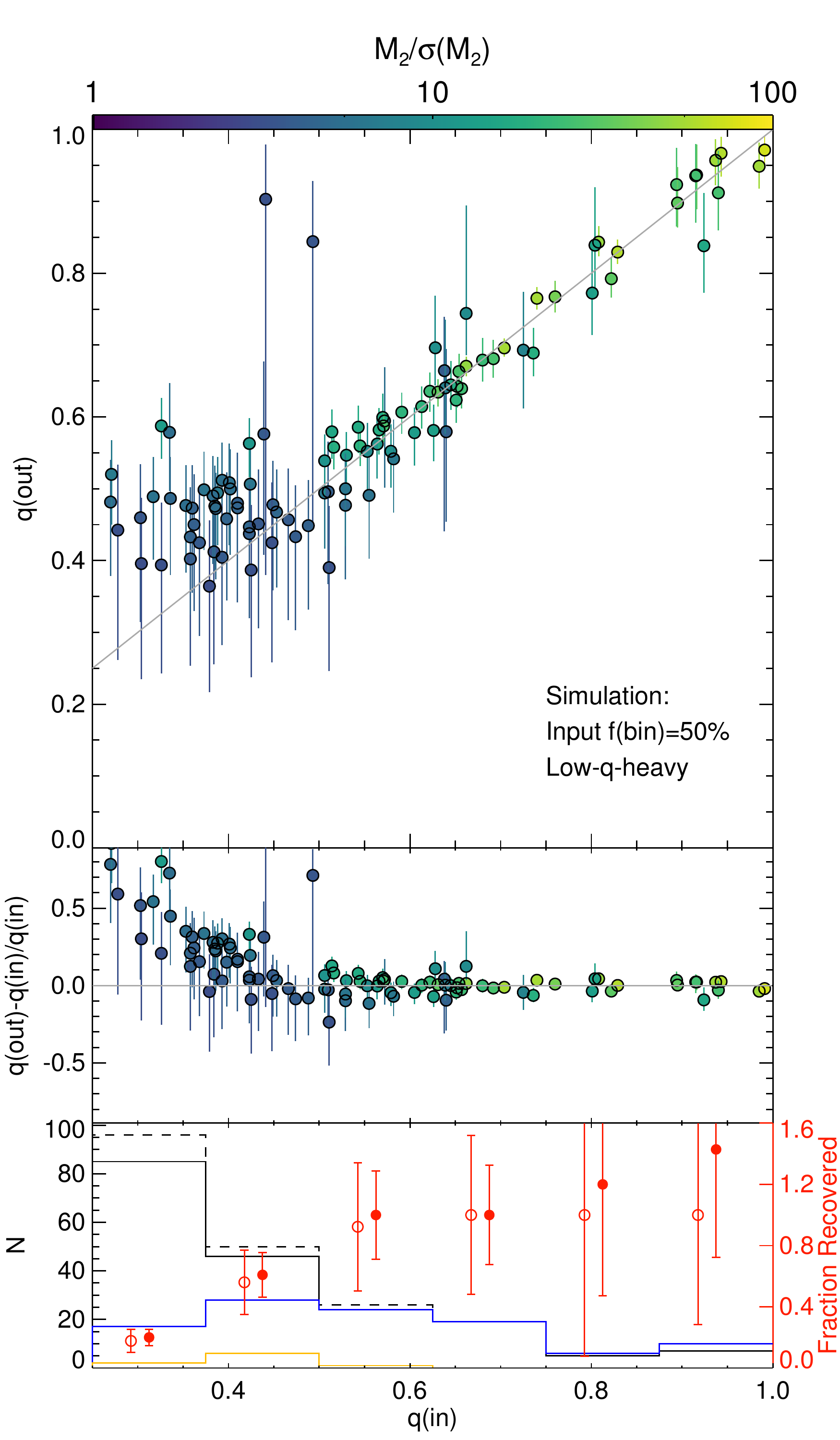}{0.33\textwidth}{}
		  \fig{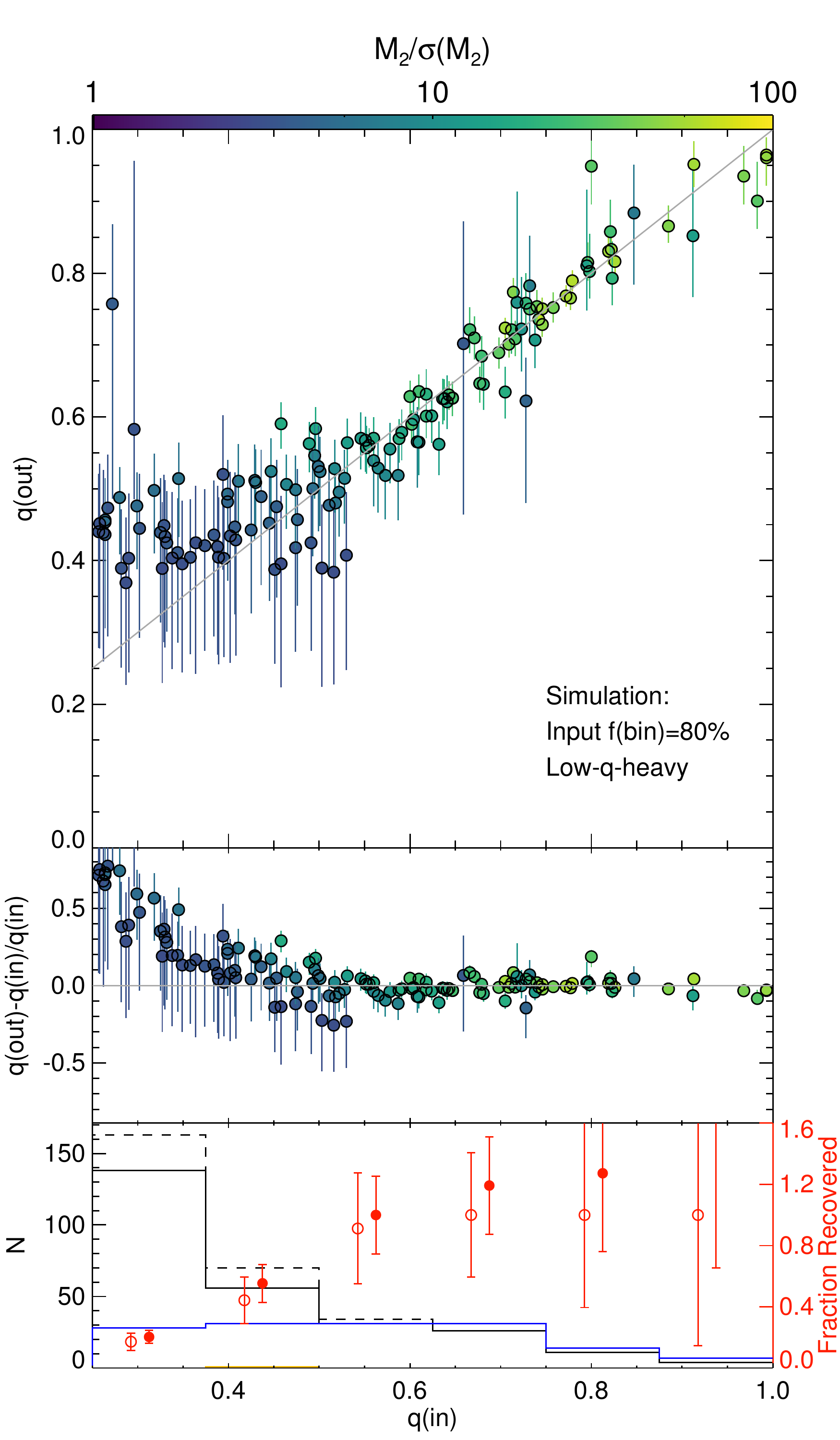}{0.33\textwidth}{}
		  \fig{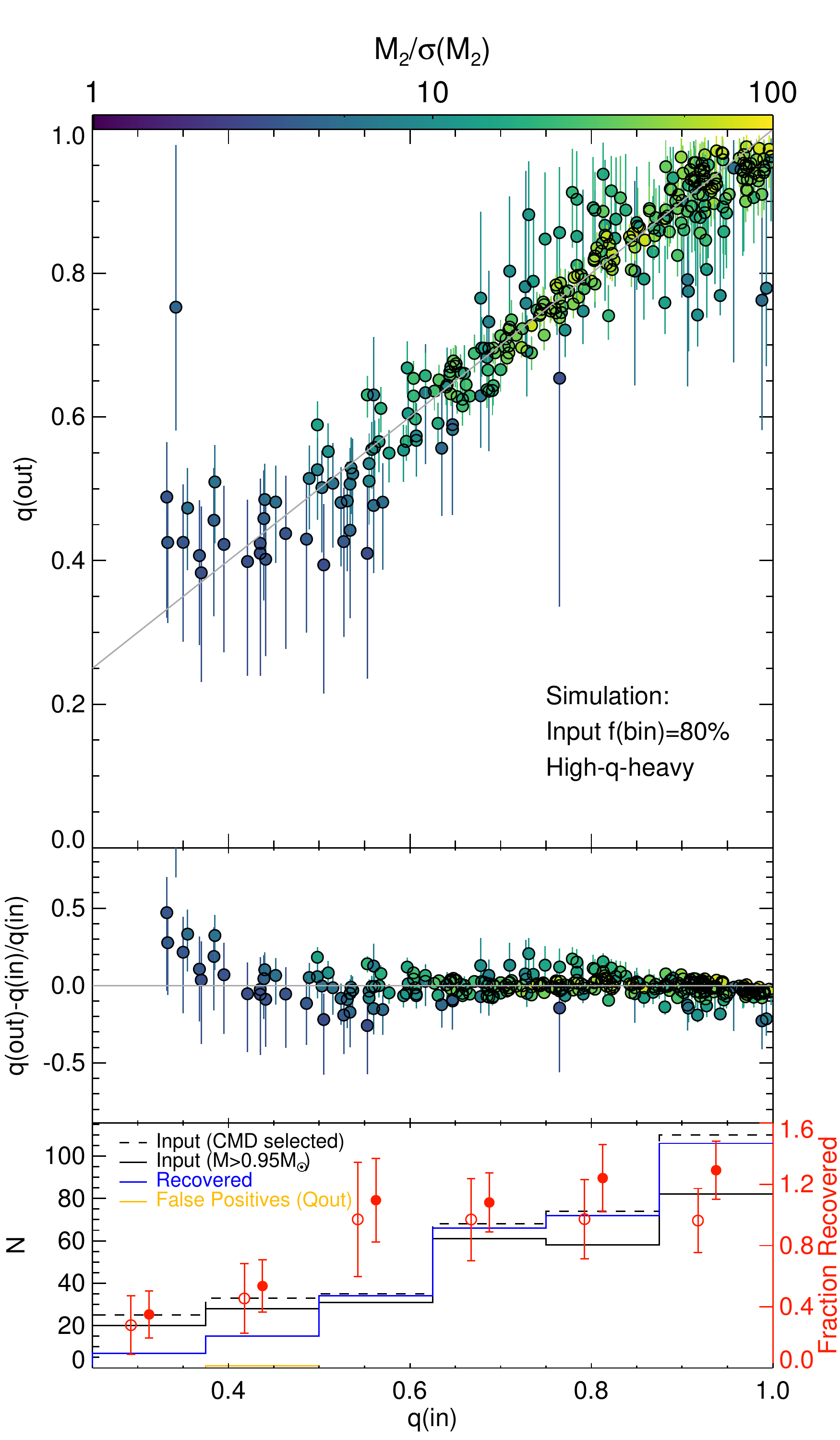}{0.33\textwidth}{}}
\caption{For main sequence binaries from the three example cases shown in Fig.~\ref{simcmdoutfig}, we plot recovered versus input mass ratio $q$ (top) and its fractional residuals (middle), color coded by output $M_{2} / \sigma (M_{2})$.  In the bottom panel for each example case, we show histograms of the input mass ratio $q(in)$ drawn from the primary-mass-limited sample ($M > 0.95M_{\sun}$) using solid black lines, while the input mass ratio distribution from the CMD-selected sample is shown using dotted black lines, illustrating that while the CMD-selected sample is larger, the \textit{distributions} of $q(in)$ are similar.  The $q(in) $ distribution for the subset of the CMD-selected sample which was successfully recovered is shown in blue, and a histogram of the \textit{output} mass ratios of input single stars misidentified as binaries (i.e.~false positives) is shown in orange, clustered near $q(out) \sim$0.4 regardless of binary fraction or mass ratio distribution.  The fraction of input binaries which was successfully recovered 
as a function of $q(in)$ is overplotted in red with Poissonian error bars, using open circles to illustrate the recovery fraction versus the CMD-selected sample (offset by $\Delta q(in)$=-0.02 for clarity), and filled circles to directly compare the ratio of recovered binaries from the CMD-selected sample to the number of input binaries from the larger primary-mass-selected sample. 
 \label{simqinoutfig}}
\end{figure}

With the goal of measuring not only the global binary fraction of NGC 188, but also constraining the \textit{distribution} of binary mass ratios, we now need to assess how well BASE-8 is able to recover $q(in)$ as a function of the observed mass ratio $q(out)$, and how the recovery fraction depends on $q(in)$, binary fraction, and mass ratio distribution.  To this end, in Fig.~\ref{simqinoutfig} we plot several quantities as a function of $q(in)$ for the three example simulations shown in Fig.~\ref{simcmdoutfig}, restricting our sample to main sequence binaries 
to enable a comparison to the spectroscopic results of \citet[][see their fig.~8]{gellerhardbin}.  For each example cluster realization, the top panel of Fig.~\ref{simqinoutfig} directly compares $q(out)$ against $q(in)$, where the value of $q(out)$ for each star is the median over all iterations, and the error bars illustrate the 16th and 84th percentiles.  In the middle panel of Fig.~\ref{simqinoutfig}, we illustrate the fractional residuals, also as a function of $q(in)$.  It is apparent that the recovered mass ratio $q(out)$ deviates from $q(in)$ at both very high ($q(in) \gtrsim$0.9) and fairly low ($q(in) \lesssim$0.5) true mass ratios.  At the high-$q$ end, this is simply due to our choice to use the median and standard deviation to quantify $q(out)$, since $q$ by definition cannot be larger than one.  For this reason, the PDFs of stars with $q$ close to one will be truncated at one and will have median values lower than they would be for an identical distribution centered at smaller $q$.  Meanwhile, at decreasing $q(in)$, we can see that the statistical significance of the recovered binaries, indicated by the colorbar, decreases.  

The upper and middle panels of Fig.~\ref{simqinoutfig} illustrate how a relationship may be constructed between our observed $q(out)$ distribution and its underlying $q(in)$ distribution, but we must now account for the fact that this $q(in)$ distribution is drawn from the CMD-selected sample, whereas our ultimate goal is the recovery of the true, primary-mass-limited $q(in)$ distribution.  To this end, in the bottom panel of Fig.~\ref{simqinoutfig}, we illustrate the $q(in)$ distribution from the CMD-selected sample with a black dotted line, and the true $q(in)$ distribution from the input primary-mass-limited sample with a black solid line, in bins of 0.125$q$.  Here, we see that the two $q(in)$ distributions have a similar shape, 
yielding values of $f(rec)$ close to 
one.  However, the total \textit{number} of input binaries in the CMD-selected sample is larger, illustrating that the \textit{number} of $M_{1} <$0.95$M_{\sun}$ binaries is larger than the number of binaries in the BSS region excluded by the CMD cuts plus those \textquotedblleft hidden\textquotedblright{} in the turnoff.  This is illustrated using the open and filled red symbols, which compare, as a function of $q(in)$, the number of binaries recovered from the CMD-selected sample (shown in blue) to the number of input binaries in the CMD-selected sample (shown as open red circles), and also to the (smaller) number of input binaries in the primary-mass-limited sample, shown as filled red circles.  In other words, while the CMD-selected and primary-mass selected samples give similar binary \textit{fractions}, the primary-mass-selected sample has a smaller \textit{number} of binaries, giving completeness fractions larger than one when directly compared to a larger CMD-selected output sample.

\begin{figure}
 \gridline{\fig{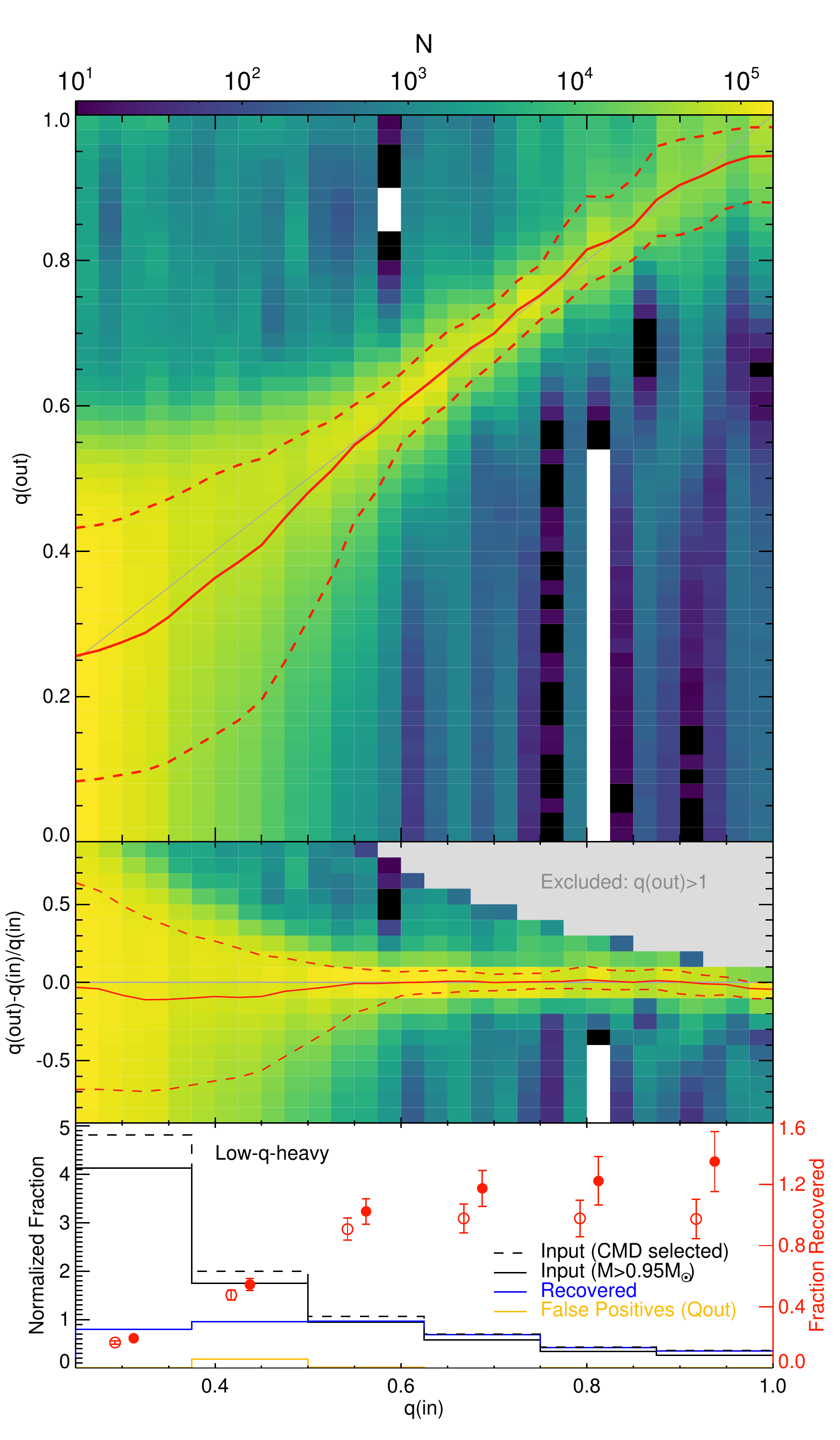}{0.5\textwidth}{}
           \fig{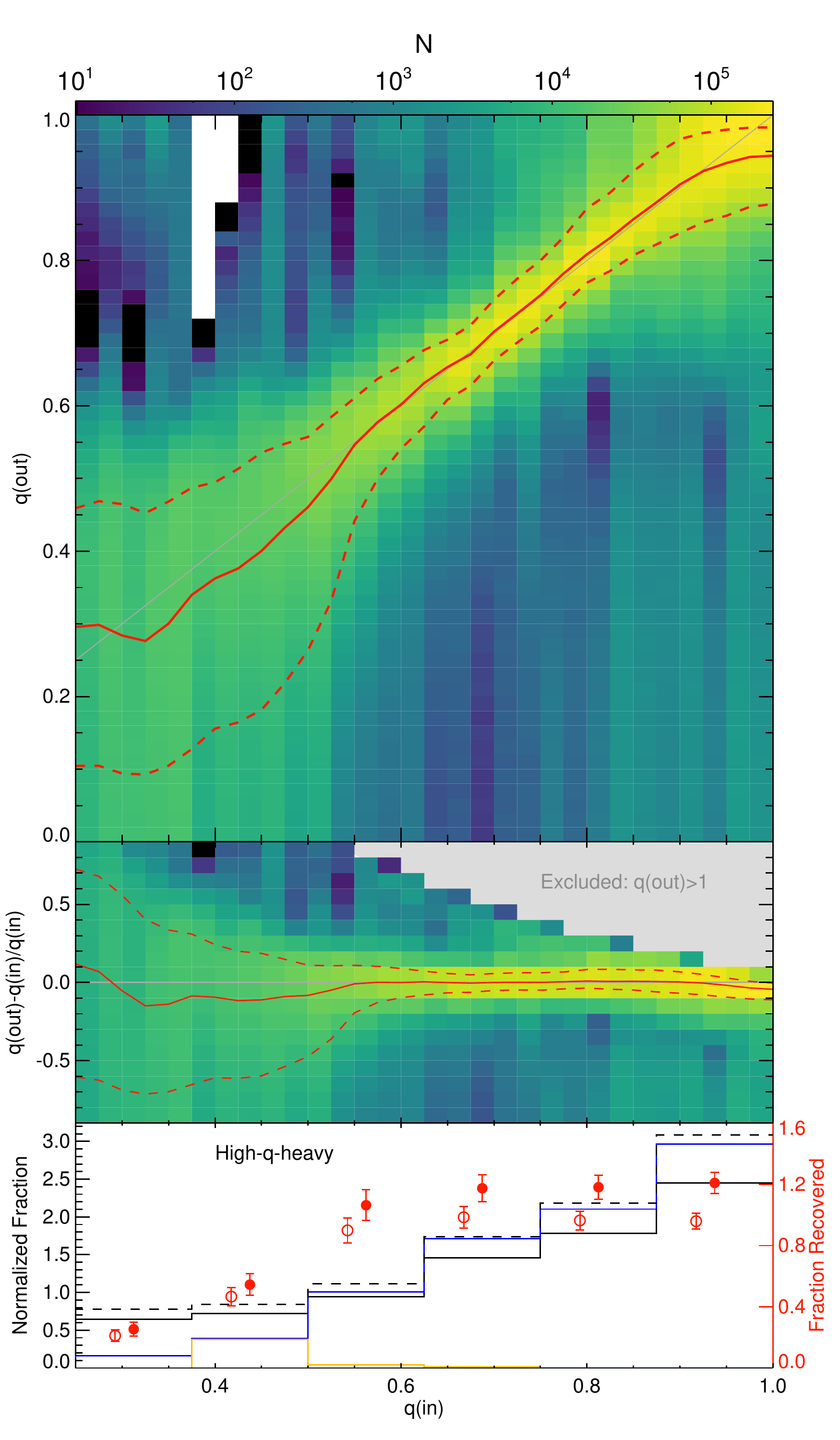}{0.5\textwidth}{}}
\caption{As in Fig.~\ref{simqinoutfig}, but showing results from all MCMC iterations for all simulation runs concatenated together for the low-q-heavy (left) and high-q-heavy (right) mass ratio distributions.  The red solid and dashed lines indicate the running median and 16th-84th fractiles respectively.  The simulations indicate that both the accuracy and precision of recovered mass ratios suffer at low ($q <$0.5) mass ratios, while for $q >$0.9,input mass ratios are recovered to better than 10\% at 1$\sigma$ but are biased low due to output mass ratio distributions being truncated at an upper limit of $q=$1.    
 \label{simallqinoutfig}}
\end{figure}

In order to use the simulations to translate the observed \textit{output} mass ratio to the actual (i.e. \textit{input}) mass ratio, we concatenate multiple simulation runs with different assumed input binary fractions, minimizing the impact of small number statistics, and plot the results in Fig.~\ref{simallqinoutfig} in the same format as in Fig.~\ref{simqinoutfig}.    However, in Fig.~\ref{simallqinoutfig}, rather than showing results for individual stars, we now exploit the individual posterior values available over all MCMC iterations of all simulations with the same input mass ratio distribution.  

For each of the two input mass ratio distributions summarized in Fig.~\ref{simallqinoutfig}, the red solid and dashed lines in the upper two panels illustrate the running median and standard deviation (16th-84th percentile) as a function of $q(in)$, highlighting the systematically underestimated values of $q(out)$ for $q(in)\gtrsim$0.9, and the loss of precision \textit{and} accuracy for $q(out)\lesssim$0.5.  The simulations, which were assigned photometric errors so as to mimic our observed catalog, show that for higher mass ratios, $q >$0.5, we recover binaries and their mass ratios quite well.  
However, below $q$=0.5, the fraction of recovered binaries drops sharply, and of the minority which are recovered, their mass ratios tend to be underestimated, more severely for more extreme mass ratios. Importantly, the similarity between the left and right plots of Fig.~\ref{simqinoutfig} illustrates that while the raw output mass ratio distribution unsurprisingly depends on the input mass ratio distribution, \textit{all} of the following quantities are insensitive not only to the binary fraction, but also to the assumed mass ratio distribution: First, the relationship between $q(out)$ versus $q(in)$, including the systematic offsets described above.  Second, the output mass ratios $q(out)$ of false positives, which peak at $q(out)<$0.5 in all cases.  Third, the fraction of binaries which are recovered as a function of mass ratio, which remains above 1 for high mass ratios (recall that this fraction, shown using filled red circles, is with respect to the smaller primary-mass-limited input sample) and universally drops off fairly sharply for $q(in) <$0.5. 

For this reason, we restrict our analysis of our NGC 188 catalog to stars with $q (out)>$0.5 since the simulations illustrate that this is the regime where binary properties are most reliably recovered.  Specifically, the simulations find 
that of binaries with $q(in)>$0.5, over 85\% are recovered with $q(out) >$0.5, and the vast majority ($>$80\% in all cases) of these are recovered with output mass ratios within 10\% of their input mass ratio (again, regardless of assumed mass ratio distribution).  
Furthermore, because false positives are preferentially recovered with $q(out)<$0.5, restricting our analysis to $q(out) >$0.5 reduces both their contribution to the binary sample (from over 40\% in some cases) and its dependence on binary fraction and mass ratio distribution, to 1.1$\pm$1.6\% (0.8$\pm$2.5\%) for the low-q-heavy (high-q-heavy) mass ratio distribution.  
After some experimentation, we found that adopting a stricter criterion to recover binaries such as $M_{2} / \sigma\left(M_{2}\right) >$5 or 7 can reduce the fraction of contaminants even further, but this comes at the cost of Poissonian uncertainties stemming from fewer recovered true low-mass ($q <$0.5) binaries.  Since our goal is to assess the ensemble properties of the cluster binary population with $q>$0.5, such a tradeoff does not affect our conclusions, and could be tailored to different science cases using an ensemble of simulations as we have done here.

In summary, while restricting our sample to $q(out)>$0.5 yields non-zero contamination from both false positives and true binaries with $q(in) <$0.5, the simulations illustrate that these contaminants have a fractionally small ($\lesssim$25\%) contribution which is effectively compensated by a similar fraction of unrecovered input binaries.  In addition, the simulations illustrate that when restricting our analysis to $q >$0.5, our ability to recover the true (primary-mass-limited) binary fraction is greatly improved, as the 
selection biases affecting a CMD-selected sample nearly cancel, with no significant dependence on binary fraction and a mild dependence on the assumed mass ratio distribution.  The net effect is that the $q>$0.5 binary fraction is recovered quite well, not only in an absolute sense as discussed in Sect.~\ref{simbinfracsect}, but in a relative sense as well: The MAD is 5\% (4\%) for the low-q-heavy (high-q-heavy) mass ratio distribution, which is 36\% (48\%) of the Poissonian uncertainty on the $q >$0.5 binary fraction.

\section{Results \label{resultsect}}
\subsection{Comparison to Spectroscopic Sample}
\begin{figure}
\gridline{\fig{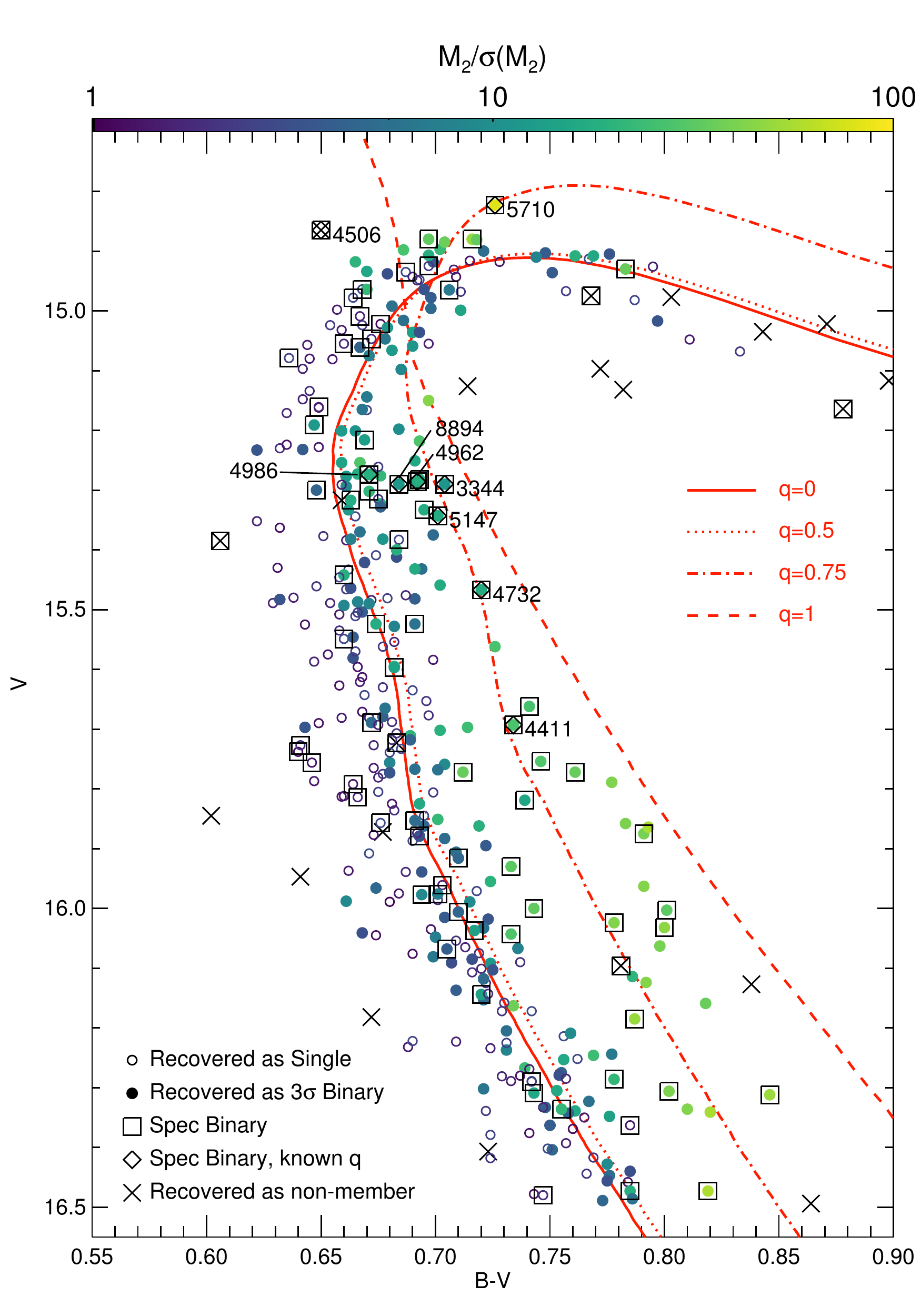}{0.5\textwidth}{}
          \fig{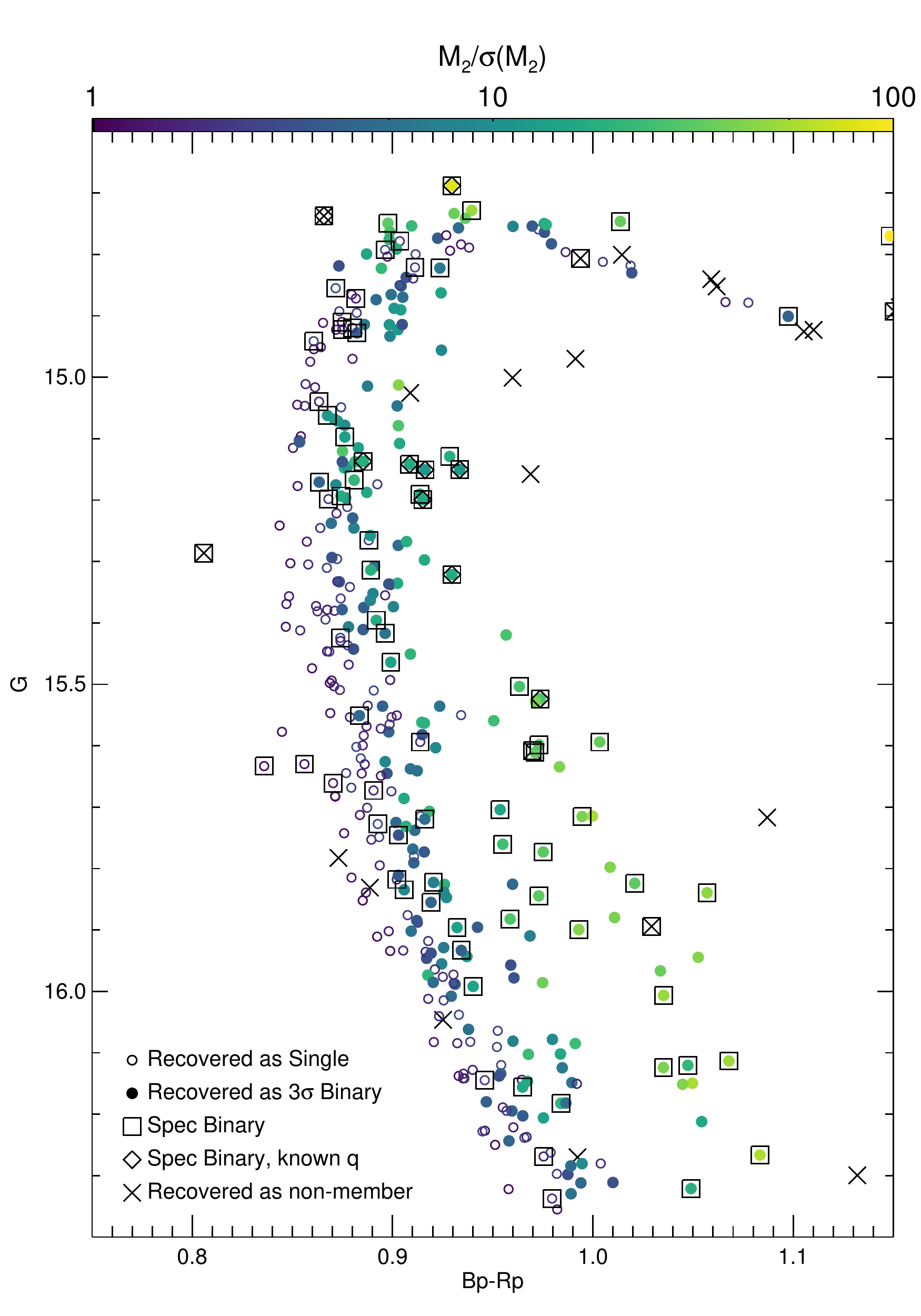}{0.5\textwidth}{}}
\caption{CMDs illustrating the results of a BASE-8 run on our observed photometric catalog of NGC 188 members, in the $(B-V),V$ plane (left) and using \textit{Gaia} DR2 photometry (right).  Stars recovered as binaries, with $M_{2} > 3\sigma\left(M_{2}\right)$ (regardless of mass ratio) are shown as filled circles, and stars not recovered as binaries, with $M_{2} < 3\sigma\left(M_{2}\right)$, are shown as small open circles.  Stars are color coded by $M_{2} / \sigma\left(M_{2}\right)$ as in Fig.~\ref{simqinoutfig}.  Known spectroscopic binaries from \citet{gellerhardbin} are indicated by boxes, and those with spectroscopic mass ratios $q(spec)$ are indicated with diamonds and labeled by their ID number in the left panel.  Crosses indicate stars which were rejected as non-members by BASE-8.  In the left panel, the best fit DSED isochrone is shown in red (solid line), and the resulting sequences corresponding to mass ratios of q=0.5,0.75 and 1 are shown as red dotted, dot-dashed and dashed lines respectively. 
\label{obscmdfig}}
\end{figure}

In Fig.~\ref{obscmdfig}, we show CMDs of the NGC 188 main sequence in $V$ versus $(B-V)$ (left panel) and photometry from \textit{Gaia} DR2 (right panel).  
Stars which are binaries according to our criterion of $M_{2} > 3\sigma\left(M_{2}\right)$ are shown using filled circles, while stars which we find to be singles ($M_{2} < 3\sigma\left(M_{2}\right)$) are shown using smaller open circles.  All stars recovered as likely ($>$50\%) cluster members have been color-coded (logarithmically) by $M_{2} / \sigma\left(M_{2}\right)$ as in Fig.~\ref{simqinoutfig}, while the stars with output membership probabilities of $<$50\% are indicated by black crosses, noting that these constitute a small (6\%$\pm$1\%) but non-zero fraction of our main-sequence sample.  

\begin{figure}
\gridline{\fig{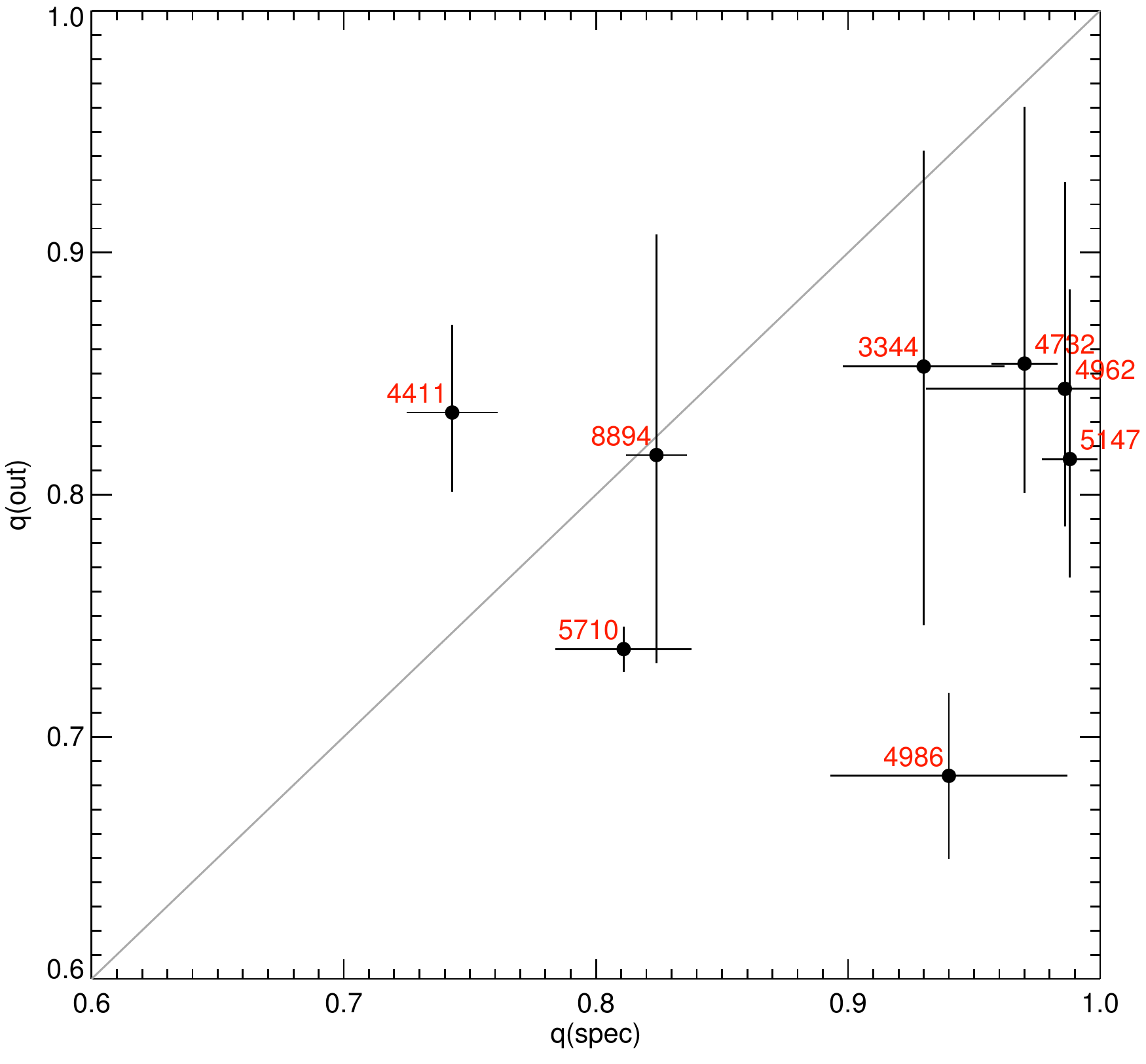}{0.4\textwidth}{}}
\gridline{\fig{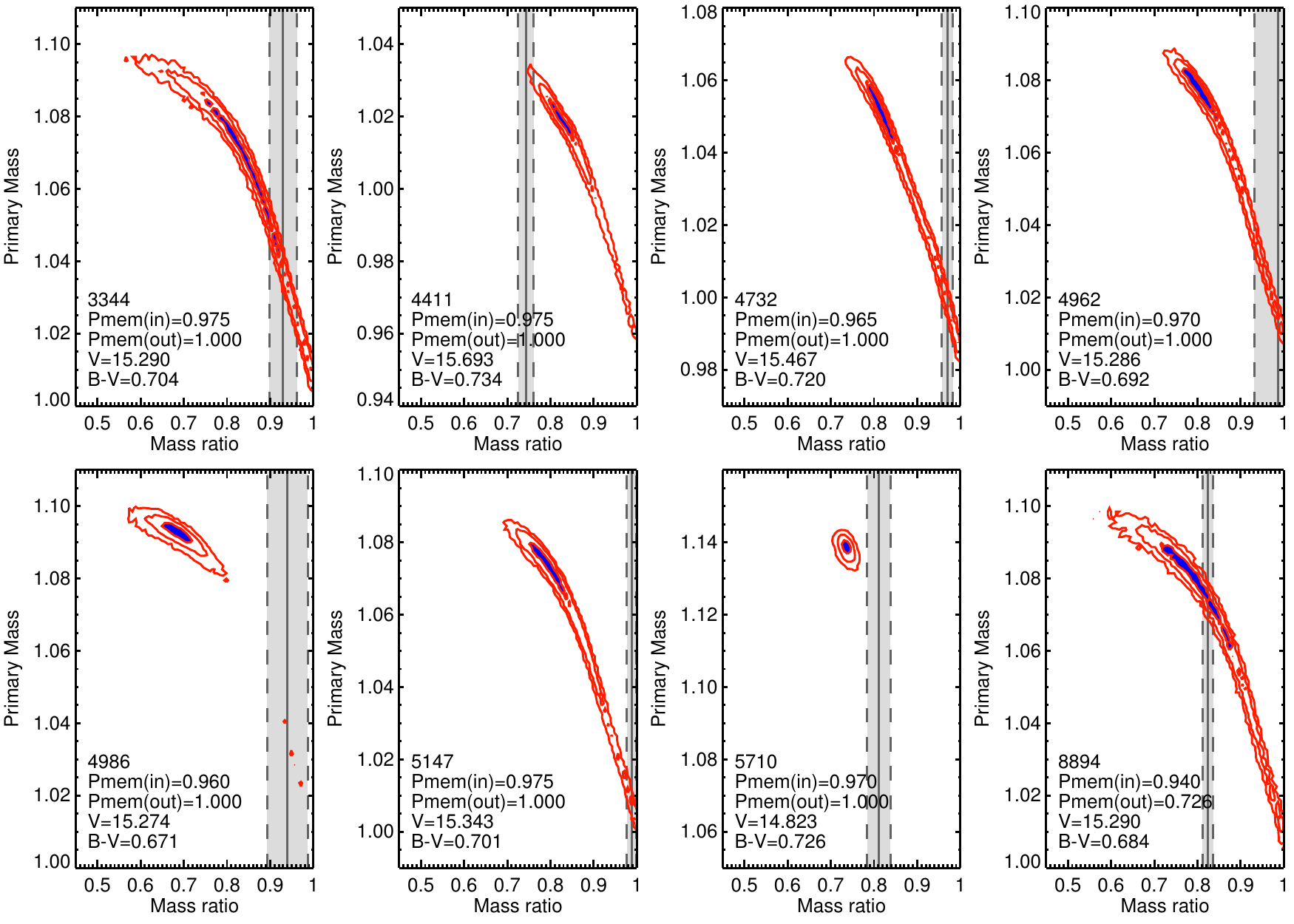}{0.95\textwidth}{}}
\caption{\textbf{Top:} For the 8 main sequence binaries with spectroscopic mass ratios from \citet{gellerhardbin}, a comparison of the mass ratios determined from our photometric catalog by BASE-8 versus the spectroscopic mass ratio.  Points represent the median and error bars represent the 16th-84th fractile interval, and stars are labeled in red by their WOCS ID number corresponding to Fig.~\ref{obscmdfig}.  \textbf{Bottom:} PDFs of primary mass versus mass ratio for each of the eight main sequence binaries with known spectroscopic mass ratios.  The spectroscopic value and its uncertainty are indicated by the grey vertical solid and dashed lines respectively, and our recovered values are shown using contours of 3, 2 and 1$\sigma$, with the latter shaded in blue.  For each star, its ID number, its input (from proper motions and/or radial velocities) and output membership probability, and its $(B-V),V$ color and magnitude are given in the lower left. The y-axis values are shifted from star to star, but the scale of both axes is kept constant in all panels.
\label{specfig}}
\end{figure}

Our choice of NGC 188 as a test case is motivated by the extensive long-term spectroscopic database available, against which we may compare the binary properties we derive photometrically.  To this end, known spectroscopic binaries from \citet{gellerhardbin} are indicated using boxes in Fig.~\ref{obscmdfig}, and the subset of these which have spectroscopic mass ratios are indicated using diamonds.  Of the latter subsample, there are nine on the main sequence, labeled by their WOCS ID \citep{platais}.  
Of these nine main-sequence binaries with spectroscopic mass ratios $q(spec)$, one (WOCS 4506) is discarded as a non-member by BASE-8, and for the remaining eight, we plot the mass ratio recovered by BASE-8 versus their spectroscopic mass ratio in the upper panel of Fig.~\ref{specfig}.  The plotted y-axis values represent the median (points) and 16th to 84th percentile range (error bars), and the diagonal grey line indicates equality.  While it appears that the majority of this subsample are recovered with mass ratios $q(out)$ discrepant with their spectroscopic mass ratio $q(spec)$ by somewhat more than 1$\sigma$, there are two biases affecting these raw values.  First, the subsample of spectroscopic binaries with measured values of $q(spec)$ is likely biased towards high mass ratios, as compared to the mass-ratio distribution of the full NGC 188 binary sample; larger $q(spec)$ values are more easily measured in general \citep{gellerhardbin}, and both stars must be nearly the same luminosity to observe both spectra and hence derive a spectroscopic mass ratio.  Second, and relatedly, in the uppermost range of $q(spec) > 0.9$, where the majority of this subsample lies, a bias in $q(out)$ operates to slightly underestimate the true value of $q$.  This bias was discussed in Sect.~\ref{simsect} and can be seen in Fig.~\ref{simallqinoutfig} for stars with $q(in) > $0.9.  Accounting for this bias would suggest an upward revision in $q(out)$ (and possibly its uncertainty) by $<$0.1, bringing the majority of the subsample into reasonable (1$\sigma$) accord with the spectroscopically measured mass ratios.  
 
At the same time, it is quite illustrative to examine the distributions of output primary mass $M_{1}$ versus output mass ratio for these eight stars, shown in the lower portion of Fig.~\ref{specfig}.  There, the two-dimensional PDFs are shown as contours at 1, 2 and 3$\sigma$, and the 1-$\sigma$ contour is filled in blue for clarity, while the spectroscopically measured mass ratio $q(spec)$ is indicated by a vertical grey line, and the shading and vertical dashed lines represents its uncertainty.  In most cases, the binaries follow a very restricted locus in $q-M_{1}$ parameter space, and while the 1-$\sigma$ uncertainties are fairly symmetric about the medians (seen in the upper panel of Fig.~\ref{specfig}) when projected in one dimension, the full distributions tend to be somewhat tailed, such that only 2/8 stars are recovered with mass ratios discrepant from their $q(spec)$ at 3-$\sigma$ confidence.  The distributions in Fig.~\ref{specfig} illustrate the utility of an MCMC approach as implemented in BASE-8 to access distributions which may be asymmetric and/or correlated in multi-dimensional parameter space.  

Because small number statistics are a limiting factor in comparing our mass ratios to spectroscopic values, we turn now to the larger sample of 83 main-sequence binaries identified spectroscopically, the vast majority of which (74/83) are single-lined and therefore have not had their mass ratios measured previously.  
We recover 54/83 (65$\pm$11\%) of these spectroscopically identified binaries, indicated by boxes in Fig.~\ref{obscmdfig}. 

\subsection{Cluster $q >$0.5 Binary Fraction \label{binfracsect}}

BASE-8 finds a raw output $q >$0.5 binary fraction of 42$\pm$4\% for NGC 188 before making any corrections for false positives or selection biases.  
As discussed in Sect.~\ref{simsect}, simulations reveal that this raw value should already reflect the true underlying $q >$0.5 binary fraction quite well, to within $<$1$\sigma$ in the mean for \textit{either} of the two simulated mass ratio distributions.  Harnessing all of our simulation results for each mass ratio distribution together, we may now correct for contaminants, including false positives, and the two competing photometric selection biases 
(recall that we found these corrections to be insensitive to input binary fraction).  
Applying these corrections and propagating their uncertainties, we arrive at a corrected solar-type (0.95$>M_{1}>$1.15$M_{\sun}$) $q >$0.5 main sequence binary fraction of 39$\pm$4\% (42$\pm$5\%) assuming a low-q-heavy (high-q-heavy) mass ratio distribution.  

Three aspects of this measurement are noteworthy: First, applying the corrections for false positives and selection biases increases the accuracy of the binary fraction at a relatively slight cost to precision, and the uncertainties remain dominated by Poisson statistics due to the size of the observational sample.  Second, these values are calculated excluding stars classified as non-members by BASE-8 based on their multi-band photometry and photometric errors, despite their inclusion in the observational sample based on ground-based proper motion and radial velocity membership (supported by \textit{Gaia} DR2).  However, this does not affect our results beyond their uncertainties, decreasing the raw $q >$0.5 binary fraction by 2\% and the corrected values by 3\%.  Third, by restricting our analysis to $q >$0.5, the resulting binary fraction is quite insensitive to the assumed mass ratio distribution used to correct for incompleteness and contaminants, and the corrected $q >$0.5 binary fractions we derive are consistent to within their uncertainties despite the assumption of two rather different forms for the mass ratio distribution used in the simulations.  However, because we have restricted our results to higher ($q >$0.5) mass ratios where the simulations indicate mass ratios 
can be recovered relatively reliably, extrapolation of our results to the \textit{full} binary fraction clearly depends sensitively on the assumed mass ratio distribution: If we assume the low-q-heavy mass ratio distribution, our result for the $q >$0.5 binary fraction would imply a \textit{global} (0$< q \leq$1) binary fraction greater than one (albeit not at high confidence; the 1$\sigma$ lower limit is $f(bin) \geq$87\%), while assuming the high-q-heavy mass ratio distribution would yield a global binary fraction of 60$\pm$12\%.  While a more detailed investigation of the mass ratio distribution is beyond the scope of this study, it bears mention that even a comparison of two qualitatively different example cases already yields some constraints.

There remain few observational measurements of the main sequence binary fraction in NGC 188 against which we may compare our results.  Among these, \citet{gellerhardbin} estimate a spectroscopic binary fraction of 29$\pm$3\% for binaries with orbital periods less than 10$^{4}$ days.  Note that, while the \citet{gellerhardbin} spectroscopic binary fraction is limited by orbital period (and not by mass ratio), conversely, our BASE-8 estimate is limited in mass ratio (but not period); therefore comparisons of these two numbers are only illustrative.  Meanwhile, our value is in good agreement with binary fractions of $\sim$40-50\% found for old (over 1 Gyr) open clusters by \citet{binocs}.

Our measurement of the binary fraction comes with the caveat that all binaries are assumed to be unresolved, although this is likely to be a valid assumption: the hard/soft boundary in NGC 188 is at a period of $P \sim$10$^{6}$ days, corresponding to $\sim$350 AU \citep{gellernbody}.  At the distance of NGC 188 ($\sim$1.7 kpc; \citealt{ata}), a system with a physical separation of $a \sim$350 AU will be separated by $\sim$0.2$\arcsec$ on the sky, and therefore is unlikely to be resolved in our photometric catalog.  To verify whether this is the case, we measured nearest neighbor distances between each star in our input catalog and all stars in the full \citet{stetson} catalog which are not non-members (including those lacking any proper motion or velocity information) and found that no pair of candidate members is separated by $a \lesssim$6000 AU, and there were only three candidate systems separated by $a <$8000 AU.

Interestingly, the $q >$0.5 solar-type main sequence binary fraction for NGC 188 is significantly higher than the field value at high confidence: The field solar-type main sequence binary fraction for $q >$0.5 is 0.25$\pm$0.03 \citep{raghavan,tokovinin,moe}, implying that the NGC 188 value is $\sim$1.6 times higher than the field value at $>$2$\sigma$ depending on the assumed mass ratio distribution.  
Restricting the comparison to close binaries ($P <$10$^{4}$ days, $a <$10 AU), the completeness corrected solar-type close binary fraction at solar metallicity down to 0.08$M_{\sun}$ is 0.20$\pm$0.03 \citep{moe19}, placing the \citet{gellerhardbin} spectroscopic value of 0.29$\pm$0.03 for NGC 188 again higher (by a factor of $\sim$1.5$\pm$0.3), but at somewhat lower ($\sim$2$\sigma$) confidence.  Similar results are obtained when imposing the NGC 188 spectroscopic period limit ($P <$10$^{4}$ d) on the \citet{raghavan} field sample, which predicts a field main sequence binary fraction of 19$\pm$2\%.  Although a detailed dynamical investigation of NGC 188 is beyond the scope of this study, the robust detection of a significantly higher binary fraction compared to the field may be due to either the preferential loss of single stars by tidal stripping, dynamical processing of the cluster over its lifetime, and/or different initial conditions between the cluster and field binary populations.

\subsection{Radial Distribution of Main Sequence Binaries}

In Fig.~\ref{bfreqfig} we compare the radial distribution of the main sequence binaries identified with BASE-8 (shown in blue) against \textquotedblleft single\textquotedblright{} main sequence stars (i.e., those that are not identified as binaries here; again some of these stars labelled as single may in fact be binaries that are beyond our detection limit), shown using a thick black line.  A Kolmogorov-Smirnov (K-S) test comparing these two populations returns a $p$-value of {0.0657.  Thus, there is a marginal $\sim$93\% confidence that these two samples were not drawn from the same parent population, with the binary population more centrally concentrated than the single population.  This result is consistent with \citet{gellerrvs}, who performed a similar comparison between their radial velocity variables (i.e.~binaries) and non-variables (i.e.~singles; we note that the RV survey has different biases than our method, though also does not perfectly separate binaries and singles).  The K-S test comparing their two samples returned a 92\% confidence level that the binaries and singles are drawn from different parent populations, with the binaries more centrally concentrated than the singles, in excellent agreement with our results.  We have thus independently confirmed the known mass segregation of the binaries using BASE-8.  To test whether our result is affected by our criterion used to select binaries, we also perform comparisons using a more stringently selected sample of output binaries, requiring that $M_{2} > 5(\sigma M_{2})$, and find $p$=0.023, indicating that these binaries have a radial distribution (shown in green in Fig.~\ref{bfreqfig}) which differs from that of the singles at higher confidence.  Similarly, if we instead select binaries using a cut on output mass ratio, the sample with $q(out) >$0.5 (shown in magenta in Fig.~\ref{bfreqfig}) gives a $p$-value of 0.0473, while a comparison between the radial distributions of our output singles and spectroscopically-identified main sequence binaries from \citet{gellerhardbin} (shown in brown in Fig.~\ref{bfreqfig}) gives a $p$-value of 0.0509.  Accordingly, the similarity of the binary radial distributions shown in the left panel of Fig.~\ref{bfreqfig} illustrates that the significantly more concentrated nature of the binary population is robust to the exact criteria used to select binaries.

It is clear from Fig.~\ref{simphoterrfig} that in several optical filters, the photometric errors in our input NGC 188 photometric catalog have multimodal distributions resulting from the concatenation and homogenization of multiple catalogs (described in detail by \citealt{stetson}).  Furthermore, we found that stars farther from the center of NGC 188 tend to have preferentially higher photometric errors.  This is illustrated in the right-hand panel of Fig.~\ref{bfreqfig}, where we have divided the main sequence sample into two subsamples, using the photometric error distributions in Fig.~\ref{simphoterrfig} to delineate a high error (\textquotedblleft HiErr\textquotedblright{}) subsample consisting of stars that have $V$- and $R$- band photometric errors $\sigma_{V} > $0.014 or $\sigma_{R} >$0.015, and the remainder as a low error (\textquotedblleft LoErr\textquotedblright{}) subsample.  To test whether preferentially higher photometric errors at larger distances from the center of the cluster could influence our results in the left-hand panel of Fig.~\ref{bfreqfig}, we turn to the simulations, and in each simulation run, we compare the recovered binary fraction $f(bin,out)$ from the HiErr sample to that from the LoErr sample.  We find no evidence for a significant difference between the two, and the mean difference between the recovered binary fraction $f(bin,out)$ between the HiErr and LoErr sample is -0.4$\pm$2.4\%.  In the inset in the right-hand panel of Fig.~\ref{bfreqfig} we show the probability distribution of this difference, which is essentially symmetrical about zero, and showed no significant dependence on input binary fraction or mass ratio distribution.  We therefore conclude that within the magnitude ranges sampled by our observational catalog, differences in photometric errors have no significant influence on our ability to recover binaries. 

\begin{figure}
\gridline{\fig{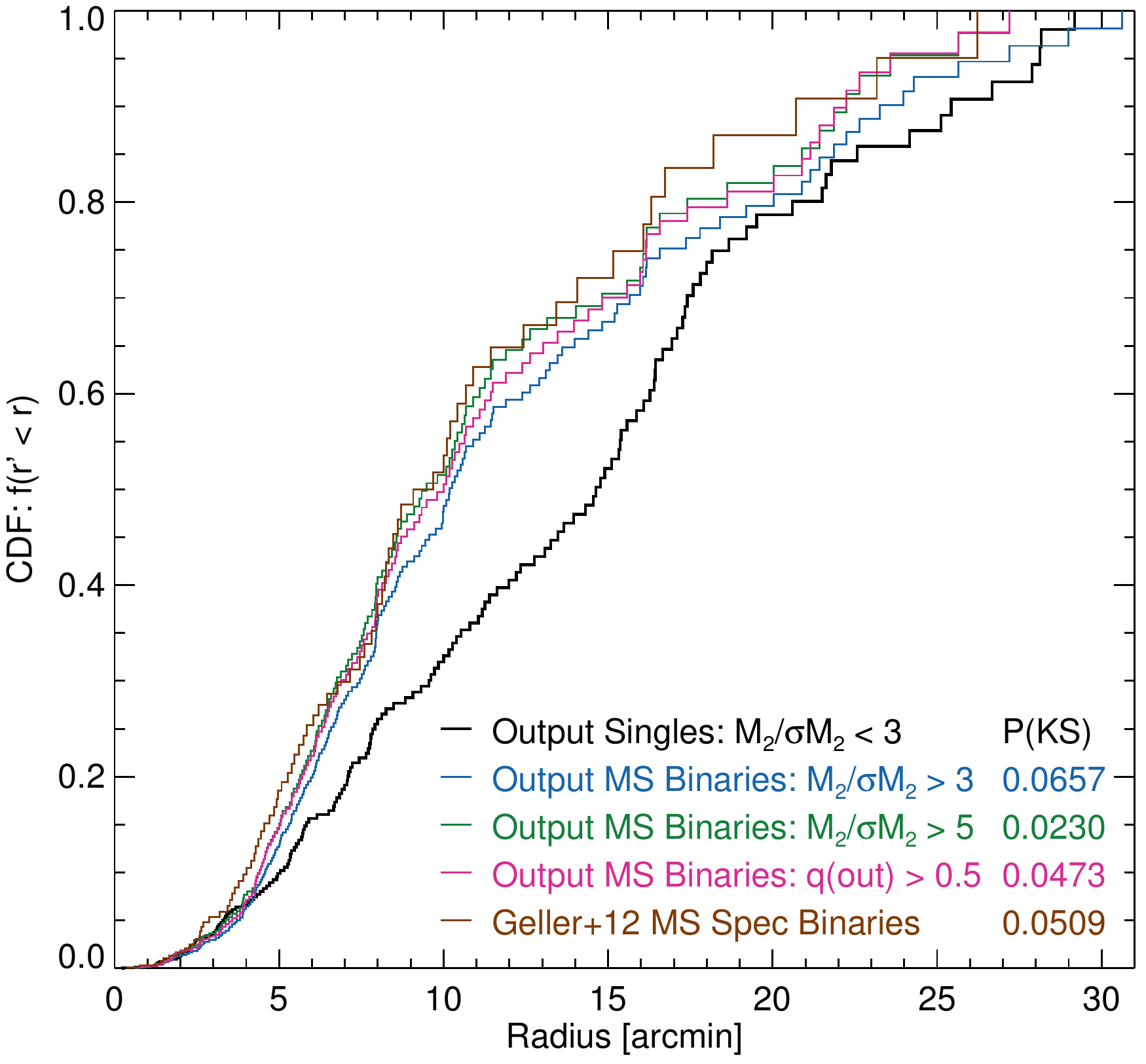}{0.49\textwidth}{}
          \fig{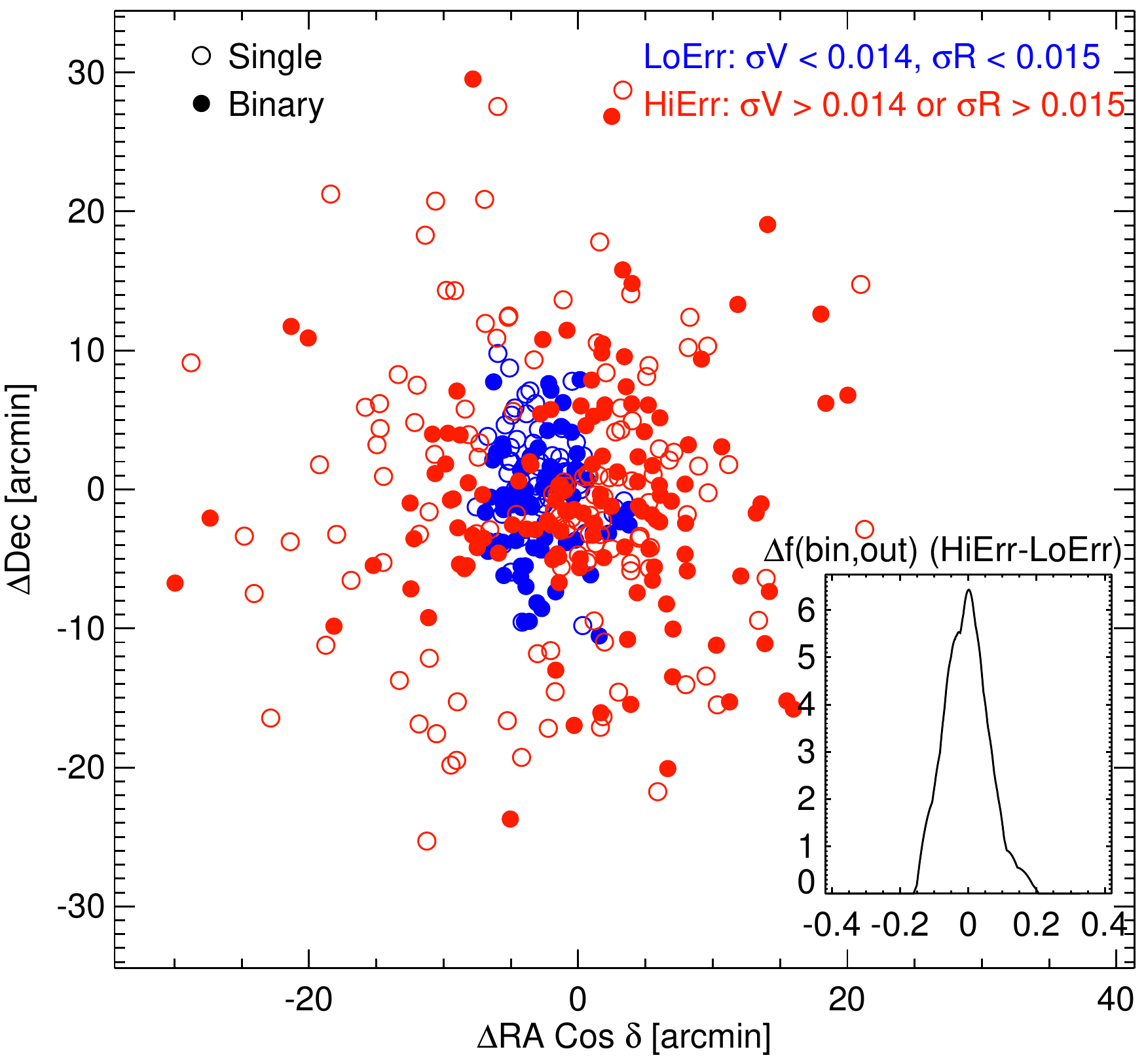}{0.49\textwidth}{}}
\caption{\textbf{(Left:)} Cumulative distribution functions comparing the radial positions of the single main sequence stars (thick black line) and binaries (blue line) in NGC 188.  The binaries are identified as described in the main text, as those objects that have $M_2 > 3\sigma \left(M_2\right)$, and all other stars are considered single.  More stringently selected binary samples, shown for comparison, have similar cumulative radial distributions to the $M_2 > 3\sigma \left(M_2\right)$ sample.  The results of a two-sided K-S test comparing the radial distributions of single stars to binaries is also given in the lower left corner for the various binary selection criteria, confirming that the ability of our method to recover the known mass segregation of binaries in NGC 188 is robust to the specifics of the binary selection criteria.  (Right:) Positions of main sequence stars in NGC 188, color coded by photometric error, revealing that stars farther from the center of the cluster tend to have higher photometric errors.  To test whether this could bias the radial distributions of recovered binaries shown in the left-hand panel, in the inset we show simulation results comparing the output binary fraction $f(bin,out)$ over all runs for stars with higher versus lower photometric errors, as given in the upper right corner. 
\label{bfreqfig}}
\end{figure}

\subsection{Binary Mass Ratio Distribution \label{massratsect}}
We have thus far discussed both the global properties of all binaries with $q >$0.5, as well as the mass ratios of individual stars for which a comparison to spectroscopic values is available.  In order to recover the mass ratio \textit{distribution}, we must correct for contaminants and incompleteness, but now as a function of mass ratio.  This process is illustrated in Fig.~\ref{obsbinfig}, employing simulation results for each of the two input mass ratio distributions.  First, a raw histogram of recovered mass ratios, shown in black, is corrected for the incidence of false positives as a function of output mass ratio, assessed from the simulations as shown in orange in the bottom panel of Fig.~\ref{simallqinoutfig}, noting that Poissonian uncertainties from both the simulations and the observations are propagated throughout, and the result is the blue histogram.  Next, the relationship between input and output mass ratio shown in the top panel of Fig.~\ref{simallqinoutfig} is used to translate the output mass ratio distribution (now cleaned statistically of false positives) to an input mass ratio distribution, shown in orange, and lastly, this input mass ratio distribution is corrected for binary detection incompleteness using the simulations, illustrated by the red points in the bottom panel of Fig.~\ref{simallqinoutfig}.  The resulting corrected mass ratio distribution is shown in red in Fig.~\ref{obsbinfig} along with its uncertainties (error bars from the intermediate steps are omitted for clarity).  

Qualitatively, the corrected mass-ratio distribution rises toward lower mass ratios for $q>0.5$, regardless of the input mass ratio distribution used in the simulations.  This general trend is also seen in the NGC 188 spectroscopic sample \citep[Fig. 8 in][]{gellerhardbin}, as well as in the field \citep{raghavan,moe,elbadry}.  In addition, all of these studies also found an excess of twins with mass ratios $q \gtrsim$0.95, which we do not see here.  However, a direct quantitative comparison between the BASE-8 and spectroscopic mass-ratio distribution is difficult given the different biases in the two techniques, while limited Poissonian statistics and the systematic offset and scatter in recovered $q(out)$ for $q(in) \gtrsim$0.9 seen in Figs.~\ref{simqinoutfig} and \ref{simallqinoutfig} may conspire to render such an excess of twins undetectable in our case.  Meanwhile, our results rule out a flat mass ratio distribution for NGC 188 at only moderate confidence, 70\% (63\%) for the low-q-heavy (high-q-heavy) simulated mass ratio distributions.

\begin{figure}
\gridline{\fig{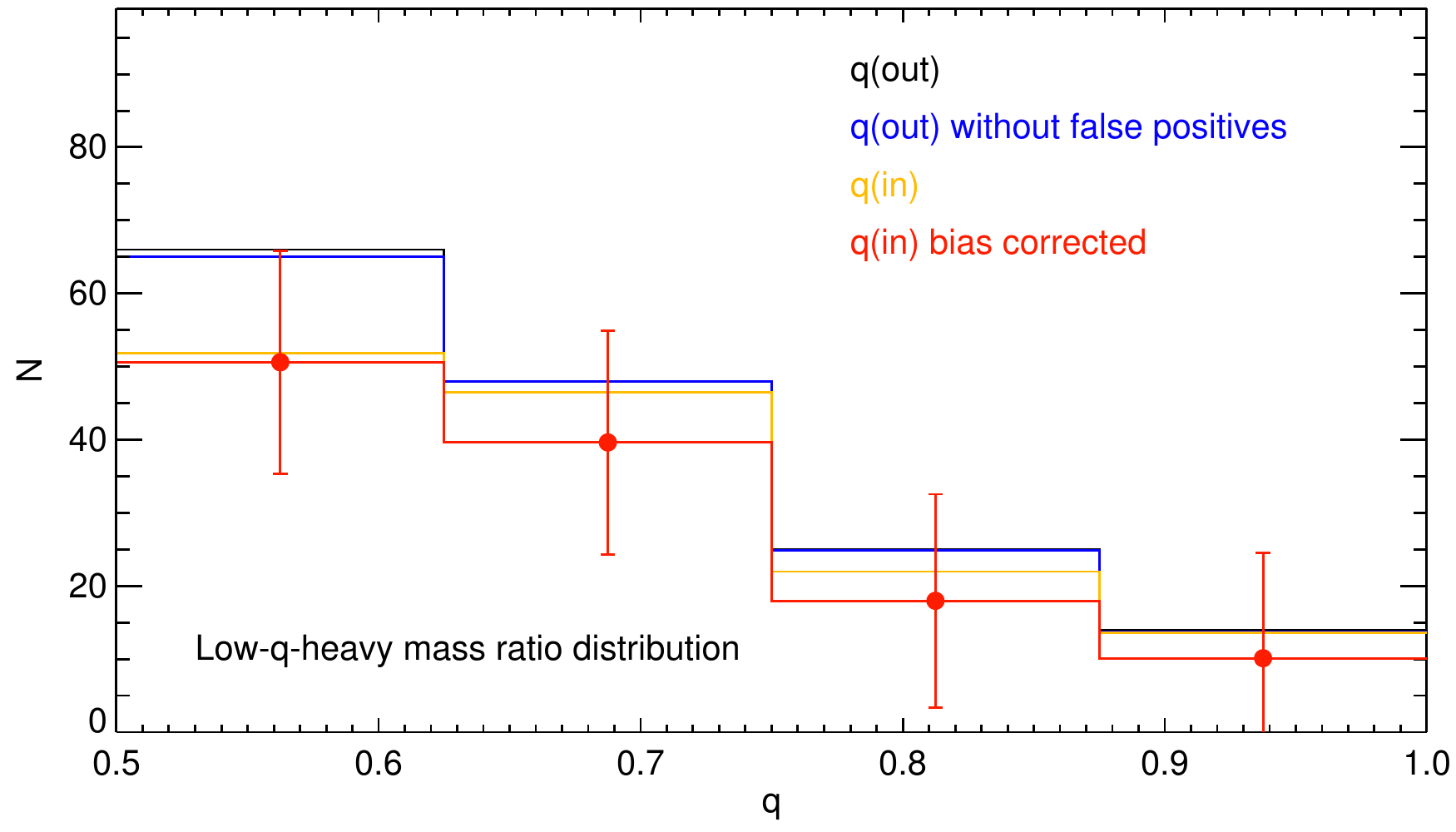}{0.5\textwidth}{}
          \fig{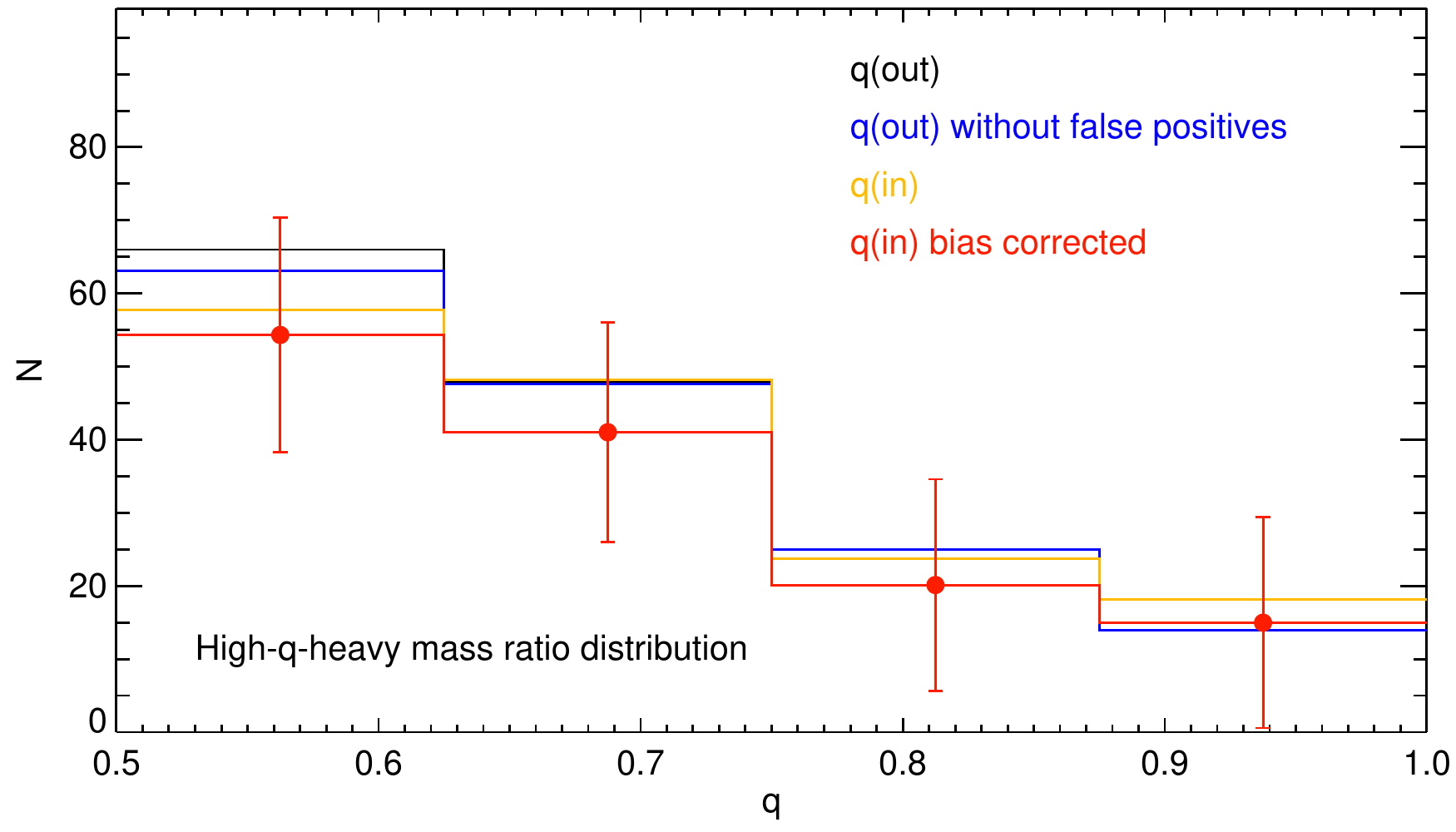}{0.5\textwidth}{}}
\caption{Raw and corrected mass ratio distributions for NGC 188 main sequence binaries, assuming a low-q-heavy (left) and high-q-heavy (right) mass ratio distribution.  The raw output mass ratio distribution from the NGC 188 photometric catalog is shown in black in both panels.  The results of all simulation runs together, shown in Fig.~\ref{simallqinoutfig}, are used to correct this distribution for false positives (blue), selection biases in output versus input mass ratio (orange), and finally, for detection incompleteness as a function of input mass ratio (red).  Despite the assumption of two schematically different mass ratio distributions used to calculate the corrections, the $q >$0.5 mass ratio distribution of NGC 188 consistently rises towards lower mass ratios.}
\label{obsbinfig}
\end{figure}

\section{Conclusions and Future Prospects}

We have used the publicly available tool BASE-8 to analyze multi-band photometry of NGC 188, with the goal of assessing our ability to both recover spectroscopically identified binaries (and their properties) and identify additional candidates.  We use simulations to quantify our ability to recover input binaries and the properties of the main sequence binary population and find the following:
\begin{enumerate}
\item{Simulations with photometric errors and a total number of stars tailored to mimic our observed catalog indicate that our ability to recover both the input binary fraction and the mass ratios of individual binaries is greatly improved for mass ratios of $q\gtrsim$0.5.  In this range, simulations indicate that we recover the input binary fraction to within 7\% in the mean.}  

\item{The simulations also reveal that for binaries with $q >$0.5, the relationship between input and output mass ratio, the contamination fraction, and binary detection incompleteness are all essentially independent of the cluster binary fraction \textit{and} mass ratio distribution.} 

\item{We recover 65\%$\pm$11\% of spectroscopically identified (single- or double-lined) cluster binaries, and we recover 8 of the 9 
double-lined binaries with spectroscopic mass ratios.}  

\item{We find a raw main sequence binary fraction of 42\%$\pm$4\% for $q>$0.5.  Correcting for systematics including false positives, binary detection incompleteness and photometric selection biases, the corrected main-sequence binary fraction for $q>0.5$ is unaffected by assumptions on the form of the mass ratio distribution to within its uncertainty, resulting in corrected values of 39$\pm$4\% or 42$\pm$5\% for a low-q-heavy or high-q-heavy assumed mass ratio distribution respectively.  This value is in reasonable agreement with recent studies employing either long-term spectroscopic campaigns or SED fitting, given their uncertainties and limitations, but significantly higher than the field solar-type main sequence binary fraction.}

\item{We derive a mass-ratio distribution that increases toward lower mass ratios within our $q>0.5$ analysis domain, and this distribution is also robust within its (Poisson-dominated) uncertainties to assumptions on the underlying mass ratio distribution used to correct for contaminants and detection incompleteness.}

\end{enumerate}

We have intentionally avoided making use of any information from \textit{Gaia} DR2 (proper motions, photometry or parallaxes) in this study so that it may serve as an independent check on our results.  Indeed, the right panel of Fig.~\ref{obscmdfig} illustrates that high-quality \textit{Gaia} photometry supports the binary nature of the candidates we have identified, while the minority of spectroscopically-identified binaries we have failed to recover using 8-band photometry lie very close to or blueward of the main sequence in the \textit{Gaia} passbands as well.  Given the accord between our results and a \textit{Gaia} proper-motion-selected sample, in future papers, we aim to extend our technique to additional clusters leveraging the \textit{Gaia} data, to place quantitative constraints on several heretofore poorly-known (or assumed) aspects of cluster binary populations, including correlation in the binary fraction and the mass-ratio distribution, with e.g., radial location within the cluster and/or cluster-wide properties such as age, size, and metallicity.

\acknowledgements

It is a pleasure to thank the referee, Maxwell Moe, for thorough and insightful comments.  This material is based upon work supported by the National Science Foundation under Grant No.\ AST-1715718.

\end{document}